\documentclass[aps,pra,reprint,superscriptaddress]{revtex4-1}
\usepackage[utf8]{inputenc}
\usepackage{color}
\usepackage{graphicx}
\usepackage{enumitem}
\usepackage{amssymb,amsmath,amsthm,amsfonts}
\usepackage{mathtools,braket,array}
\usepackage{color}
\usepackage{tabularx}
\usepackage{lipsum}
\usepackage{physics}
\usepackage{qcircuit}
\usepackage{multirow}
\usepackage{booktabs}
\usepackage{tikz}
\usepackage{float}

\usetikzlibrary{shapes.geometric, arrows}

\begin{document}

\title{Arbitrary parallel entangling gates with independent calibration \\ on a trapped ion quantum computer}

\author{Matthew Diaz}

\affiliation{Joint Quantum Institute, University of Maryland, College Park, MD, 20740 USA}
\affiliation{National Quantum Laboratory (QLab), University of Maryland, College Park, MD 20740 USA}
\affiliation{Department of Physics, University of Maryland, College Park, MD, 20740 USA}

\author{Masoud Mohammadi-Arzanagh}
\affiliation{Joint Quantum Institute, University of Maryland, College Park, MD, 20740 USA}
\affiliation{Department of Physics, University of Maryland, College Park, MD, 20740 USA}

\author{Yingyue Zhu}
\affiliation{Joint Quantum Institute, University of Maryland, College Park, MD, 20740 USA}
\affiliation{Department of Physics, University of Maryland, College Park, MD, 20740 USA}

\author{Mohammad Hafezi}
\affiliation{Joint Quantum Institute, University of Maryland, College Park, MD, 20740 USA}
\affiliation{Department of Physics, University of Maryland, College Park, MD, 20740 USA}

\author{Norbert M. Linke}
\affiliation{Joint Quantum Institute, University of Maryland, College Park, MD, 20740 USA}
\affiliation{National Quantum Laboratory (QLab), University of Maryland, College Park, MD 20740 USA}
\affiliation{Department of Physics, University of Maryland, College Park, MD, 20740 USA}
\affiliation{Duke Quantum Center and Department of Physics, Duke University, Durham, NC 27701, USA}

\author{Alaina M. Green}
\affiliation{Joint Quantum Institute, University of Maryland, College Park, MD, 20740 USA}
\affiliation{National Quantum Laboratory (QLab), University of Maryland, College Park, MD 20740 USA}
\affiliation{Department of Physics, University of Maryland, College Park, MD, 20740 USA}

\author{Arthur Y. Nam}
\affiliation{Department of Physics, University of Maryland, College Park, MD, 20740 USA}
\affiliation{ApexQuantum Inc., Ellicott City, MD, 21042 USA}

\date{\today}

\begin{abstract}
Parallel processing of information plays a critical role in accelerating computation. This includes quantum computers, where parallel processing of quantum information will play a critical role in practical quantum advantage. Here, we demonstrate a new type of parallel entangling gates in a trapped-ion quantum computer, that simultaneously provides efficient gate-pulse synthesis and calibration, as well as graph-pattern-agnostic implementation. We demonstrate the resulting reduced execution time in three well-known algorithms, exhibiting disjoint gates, a star graph and a ring graph respectively. For disjoint qubit pairs the execution time of our parallel gates is comparable to that of a single-pair entangling gate resulting in an approximately linear speed up. For all graph patterns our parallel gate fidelities are comparable to the fidelity of a single-pair entangling gate. These advantages motivate architectures featuring multiple medium length ion chains in future quantum computing devices.
\end{abstract}

\maketitle

\section{Introduction}

To bring about practical quantum advantages in the nearest future minimizing quantum gate operation times is essential and executing gates in parallel is a straightforward way to do this. Compared to a parallel execution, a series execution of computational gates can lead to an exponentially larger (or more) circuit depth or execution time, as shown in Table~\ref{tab:Depth}. These examples of often-used quantum subroutines enjoy significantly reduced execution time over a series approach, using parallel gates. It is therefore not surprising that there have been numerous works reported in the literature that demonstrate parallel entangling gates~\cite{figgatt2019parallel,grzesiak2020efficient,lu2019global,evered2023high,levine2019parallel,arute2019quantum}. 

The ability to implement a variety of patterns of entangling gates in parallel is critical to achieving the dramatic circuit depth reductions reported in Table~\ref{tab:Depth}. Trapped-ion quantum computers with a modest-sized chain of ions are uniquely suited to parallel implementation of {\it arbitrary} patterns of two-qubit entangling gates~\cite{grzesiak2020efficient}, exploiting its complete qubit connectivity graph available at all times~\cite{debnath2016demonstration}. Quantum computers equipped with parallel entangling gates whose patterns are too limited, for example due to restricted qubit connectivities (native~\cite{arute2019quantum} or engineered~\cite{evered2023high,levine2019parallel}) or gate schemes chosen (global gates~\cite{lu2019global}), suffer from suboptimal circuit depth reductions.

Prior methods of implementing in parallel an arbitrary pattern of two-qubit entangling gates in a trapped-ion quantum computer, such as \cite{grzesiak2020efficient}. However such strategies demand tailor-made gate pulses for each of the gate patterns to be parallelized, greatly increasing classical control complexity. Such complexity poses a particular problem when circuit compilation techniques are unable to adapt to restricted sets of gate patterns as is often the case. Furthermore, gate calibration in the prior methods presents a serious problem, as for $N$ qubits there are only $O(N)$ pulse amplitudes for $O(N)$ ions, i.e., one amplitude per ion, available for adjustment to calibrate for $O(N^2)$ entangling gate angles. 

\begin{table}
\vspace{-0.5em}
\caption{\label{tab:Depth} Circuit depth for oft-used quantum subroutines.}
\begin{tabular}{|c | c | c|}
\hline
Subroutines & Series & Parallel \\
\hline 
$N$-qubit Adder~\cite{Adder} & $O(N)$ & $O(\log(N))$ \\
$N$-qubit Clifford~\cite{bravyi2022constant} & $O(N^2)$ & $O(1)$ \\
$N$-control Toffoli~\cite{bravyi2022constant} & $O(N)$ & $O(\log^{*}(N))$ \\
$N$-qubit Permutation~\cite{grzesiak2022efficient} & $O(N)$ & $O(1)$ \\
\hline
\end{tabular}
\vspace{-2em}
\end{table}

In this paper, we present methods to design gate pulses for parallel entangling gates in a trapped-ion quantum computer that in principle do not require resynthesis for different gate patterns and can be individually and independently calibrated. In particular, we show $O(N^2)$ gate pulses are sufficient for the $2^{O(N^2)}$ two-qubit entangling gate patterns, potentially significantly improving the up-time of trapped-ion quantum computers which is strongly limited by, for example, gate calibration and pulse synthesis time even at a modest-sized number of qubits~\cite{maksymov2022detecting}. We achieve this by effectively separating the gate pulses into different frequency bands, grossly minimizing the crosstalk terms between them, and fully suppress the remaining crosstalk between them using a simple iterative method. The pulse duration and power requirements are comparable to those of a single entangling gate when parallelizing $O(N)$ gates. Further our pulses admit the use of tunable stability constraints as in~\cite{blumel2021efficient}. We implement the synthesized pulses on a trapped-ion quantum computer and use the parallelized gates to run several examples of quantum algorithms at significantly faster speeds, commensurate with the parallel gate patterns. 

\section{Methods}
\label{sec:Method}

 A trapped ion chain have a set of common motional modes arising from their mutual coulomb interactions and the trapping potential. Spin-motion entangling operations can be driven using blue and red sidebands on the internal spin transition. The M{\o}lmer-S{\o}rensen gate is a widely used entangling gate that uses a bichromatic light field of both blue and red sidebands to entangle the spins to the ions' shared motion. A gate solution leaves the spins entangled with each other while removing the transient spin-motion entanglement at the end~\cite{sorensen1999quantum,molmer1999multiparticle,sorensen2000entanglement}. The time-dependent M{\o}lmer-S{\o}rensen gate Hamiltonian $H(t)$ for a laser-driven trapped-ion quantum computer with $N$ qubits and $P$ modes is~\cite{blumel2021power}
\begin{align}
    H(t)=\sum_{\substack{i \in [0,N)\\ p \in [0,P)}}\eta_{ip}g_i(t)(a_p e^{-i\omega_pt})\sigma_i^x + h.c.,
    \label{eq:hamiltonian}
\end{align}
where $\eta_{ip}$ is the Lamb-Dicke parameter for qubit $i$ and mode $p$ that corresponds to the amount of participation of qubit $i$ in mode $p$. $g_i(t)$ is a pulse function such that the pulse seen by an ion is $e^{2\pi i\mu_0 t}(g_i(t){+}h.c.)$ with the single-qubit transition (carrier) frequency $\mu_0$, illuminating qubit $i$. $a_p$ is the annihilation operator for mode $p$, $\omega_p$ is the angular frequency of mode $p$, and $\sigma_i^x$ is the Pauli-$x$ matrix on qubit $i$. The gate unitary $U$ that results from applying $H$ for duration $\tau$ is then
\begin{align}
\label{eq:timeevolution}
&U(\tau)=e^{-i \left(\sum_{i,p} \eta_{ip}  (\alpha_{ip}a_p+h.c.)\sigma_i^x + \sum_{i,j,p}  \eta_{ip}\eta_{jp} \theta_{ijp} \sigma_{i}^x \sigma_{j}^x/2 \right)}, \nonumber \\
&\alpha_{ip} = \int_0^\tau g_i(t) e^{-i\omega_p t} dt, \nonumber \\
&\theta_{ijp} =  4\int_0^\tau dt_2  \int_0^{t_2}dt_1 g_{i}(t_1) g_{j}(t_2) \sin[\omega_p(t_1-t_2)].
\end{align}
To obtain parallel entangling gates, we require $\alpha_{ip} = 0$ for all $i$ and $p$ and $\chi_{ij}:= \sum_{p} \eta_{ip}\eta_{jp} \theta_{ijp}$ to be predefined values, e.g., $\pi/2$ for fully entangling gates between qubits $i$ and $j$ or $0$ for no entanglement. These $\alpha_{ip}$ and $\chi_{ij}$ constraints are achieved by tailoring the pulse functions $g_i(t)$.

Multiple methods may be used to find $g_i(t)$ that satisfy the $\alpha$ and $\chi$ constraints. The state-of-the-art method~\cite{grzesiak2020efficient} satisfies $\alpha$ constraints first, then uses an iterative approach to satisfy $\chi$ constraints. For the $\ell^{\text{th}}$ iteration  the pulses $g_{\ell'}(t)$, for $\ell'{<}\ell$, are assumed to have been solved. Successively, $\chi_{\ell\ell'}$ constraints are satisfied by shaping $g_{\ell}(t)$. 
Our method also takes an iterative approach with critical differences:
\begin{enumerate}[leftmargin=*]
\item Our approach employs a smooth pulse throughout the pulse duration $\tau$.
\item  Our approach can stabilize the pulse solutions against drift in the mode frequencies.
\item Our approach admits a single pulse solution set for all two-qubit entangling gates, from which we can draw a subset to form any graph pattern, avoiding the need to synthesize solutions for every gate pattern (see~\footnote{In principle, the method of \cite{grzesiak2020efficient} allows some patterns without synthesizing a new gate solution. For example, for a given pattern $G(V,E)$ for which a set of pulse solutions are solved, where $G$ is a gate pattern graph defined over qubits (vertices) $V$ and entangling gate (edges) $E$, a subset $G'(V',E')$ can be implemented without re-synthesizing pulse solutions when $V'$ is a proper subset of $V$ and $E'$ is a corresponding proper subset of $E$.}).
\end{enumerate}

Finding a parallel gate solution set proceeds in three steps. 

\smallskip
\noindent {\bf Step 1: Define basis functions and constraints}-- We use pulse solution basis functions $\sin(\omega_l t)$, with $\omega_l {=} 2\pi l/\tau$, where $l = 1,...,L$ is an integer. We then constrain the solutions such that all $\alpha_{ip}=0$ by using a set of linear combinations of the basis functions, i.e., $\int_0^\tau (\sum_l a_l \sin(\omega_l t)) e^{i\omega_p t} dt {=} 0$. In other words $\hat{M}\vec{a} = 0$, where $\vec{a}$ is a vector of $a_l$ and $\hat{M}$ is a $P {\times} L$-sized matrix with elements $M_{pl}=\int_0^\tau \sin(\omega_l t) e^{i\omega_p t} dt$. Note, so long as we draw our pulses from the null-space of $\hat{M}$, the pulses are theoretically guaranteed to satisfy $\alpha_{ip} = 0$.

\smallskip
\noindent {\bf Step 2: Assign modes and order the gate pulses to synthesize}-- After choosing a desired graph pattern we determine the order in which we will solve the gate pulse solutions. We ultimately want the lowest possible power requirements across the solution set, therefore we systematically consider the coupling of each gate to each mode. Candidate gate-mode pairs are ranked as follows. We first consider matrix $\hat{R}_p = \hat{A}^T\hat{S}_p\hat{A}$ where $\hat{S}_p$ is an $L \times L$ matrix with matrix elements $(\hat{S}_p)_{mn} = \theta_{mnp}$ (see Eq.~(\ref{eq:timeevolution})) and $\hat{A}$ is an $L \times (L-P)$ matrix whose columns are the orthonormalized null-space vectors of $\hat{M}$. We consider the $\nu({=}\lceil \binom{n}{2}/P \rceil)$ largest-modulus eigenvalues $\Lambda_{p,\lambda}$ of matrix $\hat{R}_p$ where $\lambda = 1,...,\nu$. Effectively, the quantity $\eta_{ip}\eta_{jp}\Lambda_{p,\lambda}$ indicates the contribution of mode $p$ to the total entanglement angle $\chi_{ij}$ for pair $(i,j)$, for a given choice of $(p,\lambda)$ making it a good proxy value for power efficiency. The problem of assigning ion gate pair $(i,j)$ to mode-eigenvalue $(p, \lambda)$ can be cast as a maximum weight matching problem~\cite{duan2014linear}. The solution to the maximum weight matching problem gives the assignments $(i,j)$ to $(p,\lambda)$ which we then rank in ascending order of the magnitude of $\eta_{ip}\eta_{jp}\Lambda_{p,\lambda}$. This allows higher power solutions (i.e. small $\eta_{ip}\eta_{jp}\Lambda_{p,\lambda}$) to be solved with fewer constraints than lower power solutions in accordance with our iterative strategy.   

\smallskip
\noindent {\bf Step 3: Parallel pulse synthesis}-- After determining the order in which to calculate the pulse solutions we solve them iteratively, each time constraining the remaining solution space to be orthogonal to all previous solutions. Assume the $m$th iteration couples $(i_m,j_m)$ and the $m'$th iteration couples $(i_{m'},j_{m'})$ with $m'<m$. In order to eliminate unwanted qubit-qubit crosstalk we set $\vec{v}_{m}\hat{S}_{\kappa,\kappa'}\vec{v}_{m'}=0$ for all combinations of $\kappa{\in}\{i_m,j_m\}$ and $\kappa'{\in}\{i_{m'},j_{m'}\}$, with $\hat{S}_{\kappa,\kappa'} {=} \sum_{p} \eta_{\kappa p}\eta_{\kappa' p} \hat{S}_p$ and $\vec{v}_{m},\vec{v}_{m'}$ are the power optimized gate pulses for iteration $m,m'$ respectively. We accomplish this by constructing the projector operator  $\hat{Q}{=}(1{-}\sum_{m',q} \vec{w}_{m',q}\vec{w}_{m',q}^T)$ where $\vec{w}_{m',q}$ are the orthonormalized vectors of $\hat{S}_{\kappa,\kappa'}\vec{v}_{m'}$ and $q {\in} [0,4)$ denotes up to four different $\kappa,\kappa'$ combinations with $\kappa{\neq}\kappa'$ assumed. Denoting $\hat{R}^*_{p} {=} \hat{A}^T \hat{Q}^T\hat{S}_{p} \hat{Q} \hat{A}$, the desired pulse $\vec{v}_m= \hat{A}\vec{r}^*_{p,\lambda}{/}{\mathcal N}_m^{1/2}$, where $\vec{r}^*_{p,\lambda}$ is the eigenvector of $\hat{R}^*_{p}$ with the $(p,\lambda)$ values determined in Step 2. The normalization factor ${\mathcal N}_m$ captures the contribution of all modes to the entanglement of $(i,j)$ and is defined as ${\mathcal N}_m{\cdot} \chi_{ij}{=}(\hat{A}\vec{r}^*_{p,\lambda})^T(\sum_{b \in [0,P)} \eta_{ib'} \eta_{jb'} \hat{S}_{p'})(\hat{A}\vec{r}^*_{p,\lambda}$).

In Appendix~\ref{app:Method1} and Appendix~\ref{app:Method2} we detail alternative methods that satisfy the constraints. Our current method (see Appendix~\ref{app:Method3} for a complete description of our current methodology) maintains the edge against the alternatives, as discussed throughout in Appendix~\ref{app:ParaMethods}. 

{\it Remarks} -- The combination of Steps 2 and 3 allows us to strictly cancel unwanted couplings without needing physically unrealistic gate powers. With feasible laser powers off-resonant coupling to other modes is naturally suppressed allowing us to assume that qubits $(i,j)$ primarily couple to the chosen mode $p$ giving us the freedom to choose power advantageous $(i,j)$-$(p,\lambda)$. However in Step 3 residual off-resonant coupling is explicitly canceled, resulting in high-fidelity power feasible gate solutions. 

The modularity of our pulse calculation can necessitate the superposition of up to $N-1$ pulse solutions on a single qubit, naively resulting in infeasible peak powers. However, it is possible to rebalance pulse power across different ions to admit similar power requirements across all participating qubits. For example, consider a star graph where multiple two-qubit gates overlap at one qubit, such as a fan-in or fan-out gate. Denoting for simplicity the pulse power on each qubit per gate as $\Omega$, implementing the pulses without the rebalancing would result in the central qubit receiving $d\Omega$ power, where $d$ is the degree of the central qubit vertex. Since the entanglement angle $\chi_{ij} \propto \Omega^2$ is a multiplicative function of the pulse powers on qubits $i$ and $j$, pulse power may be rebalanced as $\Omega/\sqrt{d}$ for the central qubit and $\sqrt{d}\Omega$ for the remaining qubits per gate such that the overall power for each qubit for all pulses combined may be $\sqrt{d}\Omega$. Since pulse power and duration may be traded off one to one~\cite{blumel2021power}, keeping the power at $\Omega$, the pulse duration becomes only $\sqrt{d}$ times longer. 
Provided our parallel pulse solutions admit a similar power requirement as the power-optimal single two-qubit gate pulse solution used in a series approach~\cite{blumel2021efficient} -- see Sec.~\ref{sec:Disc} for more details, -- compared to a serial implementation, our parallel approach can thus obtain a speed up of $O(\sqrt{d})$.

\section{Results}
\label{sec:Results}
\begin{figure}[t!]
\includegraphics[scale=1.1]{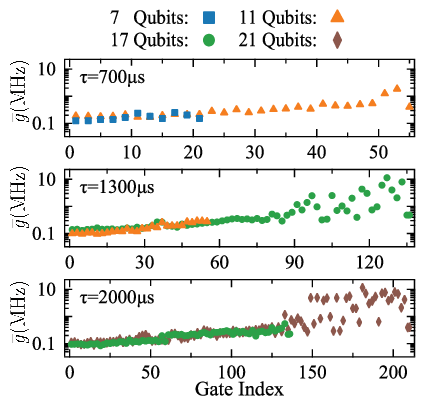}
\caption{\label{fig:PWRvsG} Calculated norm of pulse function $\overline{g}$ for parallel gate pulses as a function of gate indices, ordered according to the maximum weight matching problem solution, with the gate time specified on each plot. Fourier basis functions used span the frequency range of $(\omega_{1}/(2\pi)-100{\rm kHz},(\omega_{N}/(2\pi)+100{\rm kHz})$, where $\omega_{1}$ and $\omega_{N}$ are the lowest and highest motional mode frequencies, respectively. For visual clarity only half of the gate solutions are plotted.}
\end{figure}

Figure~\ref{fig:PWRvsG} presents the calculated power required for our parallel gates, solved for all $\binom{N}{2}$ couplings, for various gate duration $\tau$ and the number of qubits $N$. For each iteratively solved gate the power requirement is expressed as $\bar{g}$, the root-mean-square of the side-band Rabi frequency. Motional mode frequencies $\omega_p$ and Lamb-Dicke parameters $\eta_{ip}$ are generated using a theoretical model (see Appendix~\ref{app:PowerScaling} for their exact values), in line with typical trapped-ion quantum computer. We observe the pulse powers scale exponentially in the gate indices, ordered in the ascending size of $\eta_{ip}\eta_{jp}\Lambda_{p,\lambda}$. As with all known gate schemes increasing the number of qubits $N$ requires an increase in gate time $\tau$ to keep the pulse power constant as the available number of basis states drops for pulses solved towards the end. This triggers the onset of large power fluctuation for the last gate solved (see index number ${\sim}50, {\sim}80, {\sim}130$ for $\tau=700, 1300, 2000 \mu{\rm s}$, respectively).  However, the power increase and fluctuations remains relatively modest over a wide domain of the iteratively solved gates, meaning a significant portion of the entire gate set can be feasibly implemented.

\begin{figure}[t!]
\includegraphics[scale=0.55]{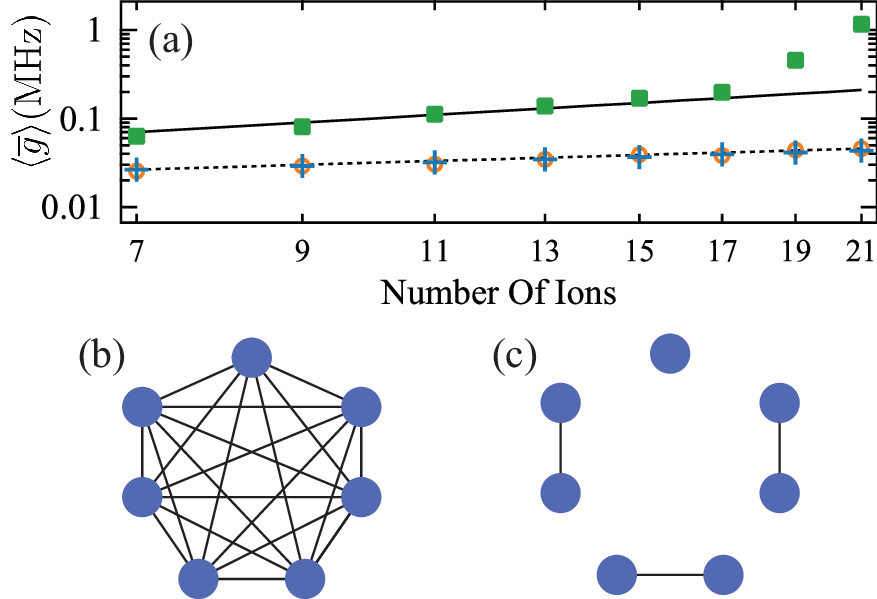}
\caption{\label{fig:PWRvsN} (a) Average calculated norm of pulse functions $\langle\overline{g}\rangle$ of parallel gate pulses as a function of the number of ions $N$ for different numbers of gates parallelized, with $\tau=2000\mu s$; Average was performed over all instances of the unique gate pattern studied. Squares: Parallelizing all $\binom{N}{2}$ gates as shown in the graph pattern example (b). Crosses: Parallelizing $\lfloor n/2 \rfloor$ disjoint gates as shown in the graph pattern example (c). Circles: Parallelizing two disjoint gates. A solid line ($\langle\bar{g}\rangle {\sim} N$) and a dashed line ($\langle\bar{g}\rangle {\sim}\sqrt{N}$) are provided to guide the eye.}
\end{figure}

Figure~\ref{fig:PWRvsN} shows the average pulse power of our parallel gates as a function of $N$, for varying number of gates parallelized, with $\tau = 2000\mu{\rm s}$. We observe that, when parallelizing all $\binom{N}{2}$ couplings, the average power scales linearly in the number of ions, whereas the power scales as ${\sim}\sqrt{N}$ when parallelizing either linear or constant number of (disjoint) gates. The onset of divergence from the linear scaling for the $\binom{N}{2}$ case at $N=19$ is due to the lack of abundant degrees of freedom (d.o.f.), consistent with the onset behaviors seen in Fig.~\ref{fig:PWRvsG}. We emphasize that parallelizing $O(N)$ gates incurs essentially no power overhead compared to parallelizing $O(1)$ gates, in the parameter space we considered. In turn, this implies $O(N)$ gates may be implemented in a similar time as a single entangling gate in that same parameter space. Note that the speed up achieved through this parallelization does not come with an increase cost in power as the $O(N)$-parallel gate pulse power is similar to that of a series gate, so long as there exists a sufficient number of d.o.f. (see also Appendix~\ref{app:PowerScaling}).

We now implement our parallel entangling gates on a trapped ion quantum computer, upgraded from the system described in~\cite{debnath2016demonstration}, configured to use ${}^{171}{\rm Yb}^+$ ions with two hyperfine ground states $|F{=}0, m_F{=}0\rangle$ and $|F{=}1, m_F{=}0\rangle$ in the ${}^{2}S_{1/2}$ manifold as the computational basis states of the qubits. We use a seven-ion chain with the middle five used as qubits. Coherent operations are implemented using Raman transitions, driven by counter-propagating beams from a mode-locked pulsed 355-nm laser, off resonantly coupling to the ${}^{2}P_{1/2}$ manifold. Our M{\o}lmer-S{\o}rensen gates ($XX$) are implemented by driving motional blue and red sideband transitions of gate pulse $g_i(t)$. Single-qubit gates are implemented by adjusting the frequency difference between the two Raman beams to match the frequency splitting between the two computational basis states. A multi-channel acousto-optic modulator is used for individual addressing and is driven by a set of arbitrary waveform generators to shape the laser pulses. 

\begin{figure*}[tp!]
\centering
\includegraphics[width=1\textwidth]{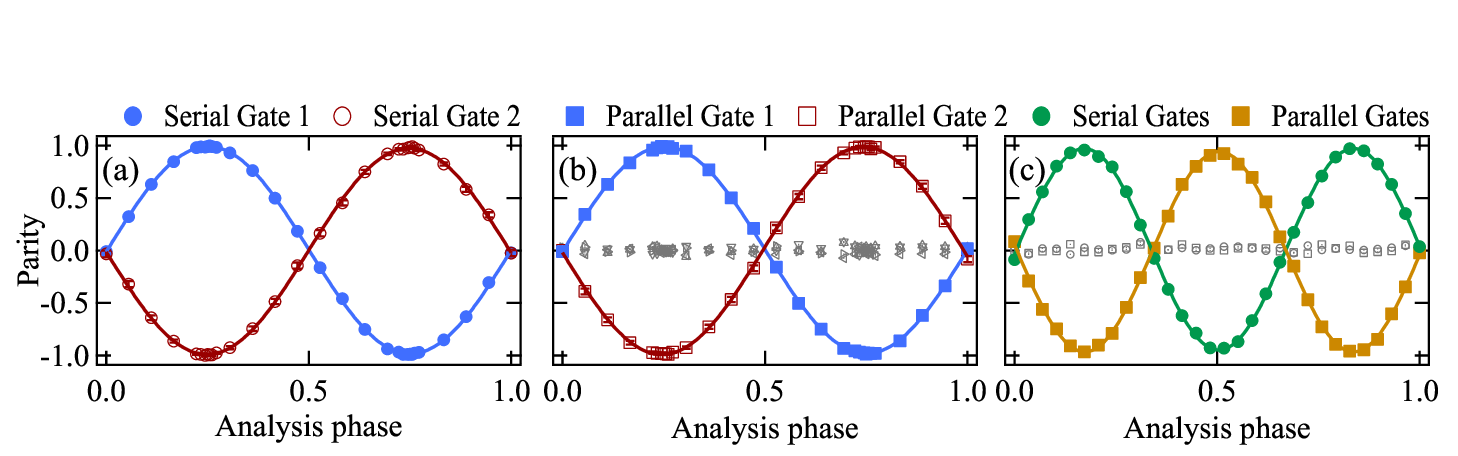} 
    \caption{\label{fig:Parity}
    Parity scans for two disjoint gates with serial implementation (a), parallel implementation (b), and two gates with one overlapping ion for both serial and parallel (c). A parity scan is conducted by first implementing the desired $XX$ gates then appending single-qubit $\frac{\pi}{2}$ rotations onto every qubit and varying the axis of rotations along the equator of the Bloch Sphere. We fit the data to $y=A\sin(\nu\pi\phi+\phi_0)$, where $A$ is the amplitude or contrast of the parity scan that we use to estimate the fidelity of the gate, $\nu=2$ for (a) and (b) and $\nu=3$ for (c), $\phi$ is the analysis phase, and $\phi_0$ is a phase offset. Grey symbols: parity between qubits over pairs not intended to be entangled. Error bars are smaller than plot symbols.}
\end{figure*}

We use our trapped ion quantum computer to compare the performance of our parallel $XX$ gates to previously established serial $XX$ gates. Figure~\ref{fig:Parity} shows the comparison of parallel vs. series for two different graph patterns, two disjoint $XX$ gates and two $XX$ gates with one overlapping ion. Specifically, we estimate the gate fidelities by using the contrast of their parity scans~\cite{Par_Fid_1,Par_Fid_2}, where the ${X}{X}$-gate sequences are followed by $\frac{\pi}{2}$ single-qubit rotations with the rotation axes scanning the equator of a Bloch sphere in unison -- Note in the overlapping ion case, several additional single-qubit gates are used to transform the state to a GHZ state, prior to the $\frac{\pi}{2}$ pulses. For the parallel gates, the estimated fidelities of the disjoint gates are $98.48(1)\%$ and $98.58(1)\%$, respectively, for the two constituent gates, while for the overlapping gates the overall fidelity is estimated $94.99(2)\%$. This may be compared to our serial gate fidelities of $98.56(1)\%$, $98.01(1)\%$, and $96.15(2)\%$, respectively.

\begin{figure*}[htp!]
\centering
    \includegraphics[width=0.9\textwidth]{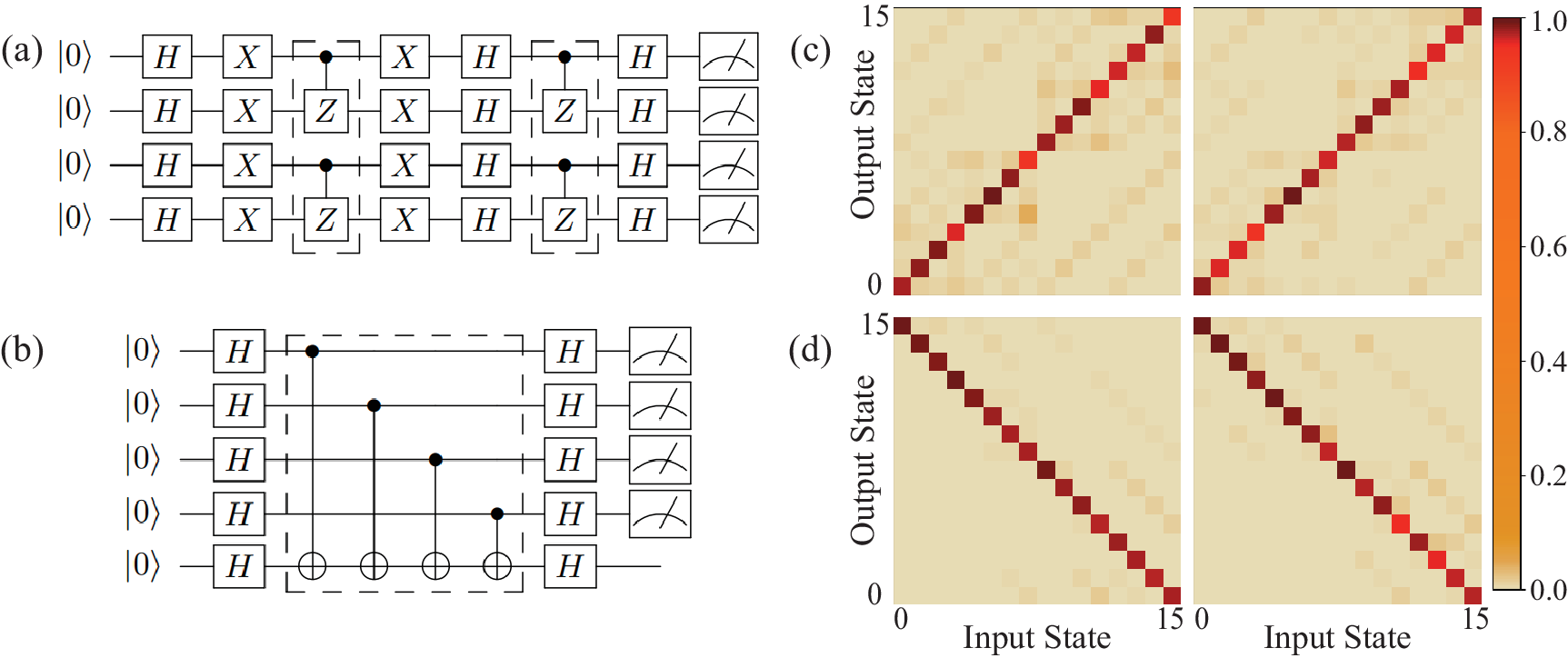}
    \caption{\label{fig:CircuitsAndResultsHSBV} Circuits and experimental results for four-qubit HS and five-qubit BV. (a)-(b): Circuits for HS, BV, respectively. The dashed boxes represent parallelized gates in our experiments. 
    (c) shows output state probabilities for the HS algorithm for serial (left) and parallel (right) implementations. (d) shows equivalent plots to (c), but for the BV algorithm. Note the color scale in non-linear for clarity.}
\end{figure*}

\begin{figure}[ht]
\centering
    \includegraphics[width=0.5\textwidth]{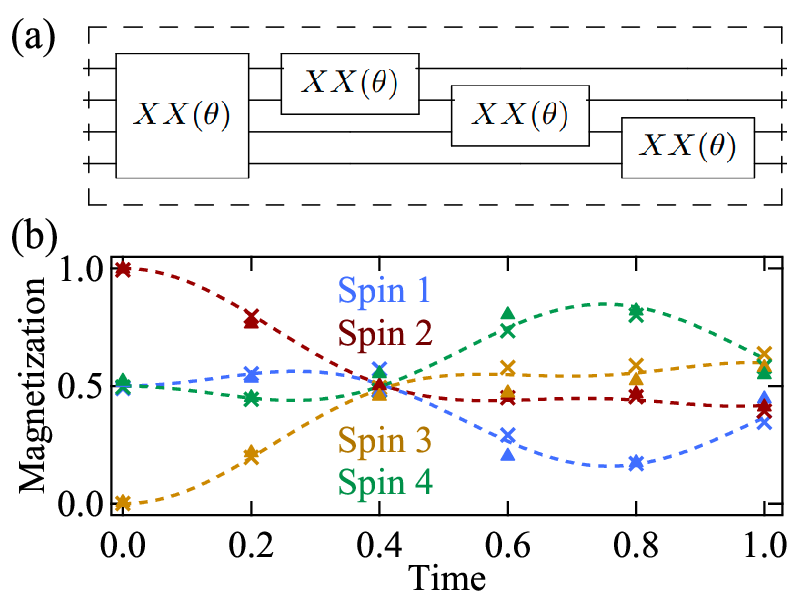}
    \caption{\label{fig:CircuitsAndResultsHH} Circuits and experimental results for four-qubit HH simulations. (a) A representative subcircuit of our HH simulation, implementing evolution corresponding to the Hamiltonian term proportional to $\sum_j \sigma_j^x\sigma_{j+1}^x$, that was parallelized (dashed box) in our experiments.
    (b) shows individual-spin magnetization for the HH, as a function of time. Crosses represent the parallel implementation, while triangles represent the serial implementation, with different colors corresponding to different spins. The dashed lines are the simulated magnetization values.}
\end{figure}

To benchmark our parallel methods in practice, we compare the parallelized and serial implementations of three algorithms: a four-qubit Hidden Shift (HS) algorithm~\cite{van2006quantum,rotteler2010quantum}, a five-qubit Bernstein-Vazirani (BV) algorithm~\cite{BV_Algo}, and a four-qubit Heisenberg Hamiltonian (HH) simulation. See Figs.~\ref{fig:CircuitsAndResultsHSBV}(a) and (b) and Fig.~\ref{fig:CircuitsAndResultsHH}(a) for the corresponding circuit diagrams. HS and BV algorithms admit straightforward verifications, i.e., their ideal output state is classical making them a clear benchmark. While the HH quantum simulation lacks these favorable benchmarking properties it is of greater general interest as it may shed light onto quantum eigenstate thermalization and many body localization~\cite{childs2018toward}, elucidating the physics behind the phase transitions between the thermalized and localized phases of magnetic systems~\cite{bonechi1992heisenberg}.

{\it HS} -- This algorithm reveals the hidden bit-string as its output state, encoded in our case as the conjugating layers of $X$ gates, as shown in Fig.~\ref{fig:CircuitsAndResultsHSBV}(a) for the 1111-bitstring example. We implement the algorithm using circuits compiled with a standard method for trapped-ion quantum computers~\cite{maslov2017basic} for all 16 oracles. Figure~\ref{fig:CircuitsAndResultsHSBV}(c) shows the measured output state probabilities as a function of the encoded oracle bit-strings, comparing parallel vs serial implementation [see dashed boxes in Fig.~\ref{fig:CircuitsAndResultsHSBV}(a)]. The average fidelity over all oracles for the parallelized implementation was $94.58(4)\%$, while for the serial implementation the average fidelity was $93.15(4)\%$, showing improvements in fidelity by parallelization. The execution time is approximately halved for the parallel implementation, as the pulse duration for entangling gates far outweigh that for single qubit gates.

{\it BV} -- As in the HS algorithm, this algorithm also reveals the encoded bit-string as its output state. The bit-string corresponds to the controls of the CNOT gates, for instance those boxed in Fig.~\ref{fig:CircuitsAndResultsHSBV}(b) for the 1111-oracle example. 
Note that in the circuit we used to implement the algorithm, shown in Fig.~\ref{fig:CircuitsAndResultsHSBV}(b), there is a fan-in CNOT, i.e., a series of CNOT gates with a shared target. We apply the power rebalancing method detailed in Sec.~\ref{sec:Method} for its parallel implementation (post standard transpiling~\cite{maslov2017basic}). While our method permits full parallelization of all four constituent gates, we limit ourselves to three gates in order to cap the maximum duration with which the phonons remain entangled to the spins. Even though four-wise parallelization reduces the total execution time from $4\tau$ to $2\tau$ in our system this causes a drop in fidelity due to $2\tau$ being comparable to our motional coherence time of less than $3\tau$. 
This results in the parallel execution time of $\tau$, $\sqrt{2}\tau$, $\sqrt{3}\tau$, and $(\sqrt{3}+1)\tau$ for oracles with one, two, three, and four 1's, which may be compared to the serial counterparts of $\tau$, $2\tau$, $3\tau$, and $4\tau$ respectively. We show in Fig.~\ref{fig:CircuitsAndResultsHSBV}(d) the measured output state probabilities as a function of the encoded oracle bit-strings. The average fidelity over all 16 oracles for the parallel implementation was $97.01(4)\%$, while for the serial implementation we obtained $98.01(4)\%$.

{\it HH} -- The system Hamiltonian we simulate is $\hat{H} = -\frac{1}{2} \sum_{j=1}^{N} \left[ J\vec{\sigma}_j\cdot \vec{\sigma}_{j+1} + h_j \sigma_j^z \right]$, where $J$ and $h_j$ characterize the strength of the interaction between neighboring spins and the magnitude of the external disorder magnetic field on spin $j$, respectively. 
Our goal is to measure the magnetization of individual spins as a function of evolution time, relevant for studying the aforementioned phase transitions.
We chose $J=\frac{5\pi}{8} \,{\rm and}\, h_j=\{0.843,-0.366,-2.047,0.003\}$ and the initial state of $|\Psi\rangle{=}-i(|0100\rangle {+} |1101\rangle){/}\sqrt{2}$, as this gives rise to a large variation in the individual spins' magnetization over the evolution time $t {\in} [0,1)$ considered. The evolution unitary is implemented using the first-order Trotter formula~\cite{suzuki1991general} and we used $\vec{\sigma}_j\cdot \vec{\sigma}_{j+1} = \sigma_j^x \sigma_{j+1}^x + \sigma_j^y \sigma_{j+1}^y + \sigma_j^z \sigma_{j+1}^z$ for Trotterization, commensurate with our native gate set [see Fig.~\ref{fig:CircuitsAndResultsHH}(a) for $\sigma_j^x\sigma_{j+1}^x$-term example]. A $\sigma^x\sigma^x$ interaction is compiled using an $XX$ gate while a $\sigma^y \sigma^y$ or $\sigma^z \sigma^z$ interaction is compiled using an $XX$ gate and single-qubit gates. Specifically, we take five trotter steps to evolve the system forward in time for $t{=}1$ and
measure the individual magnetization after each Trotter step,
whose results appear in Fig.~\ref{fig:CircuitsAndResultsHH}(b). 
We report the mean square distance between our experimental data and the results of simulation without physical error as $\bar{D}^2 {=} 0.016(3)$ in the parallel implementation and $\bar{D}^2 {=} 0.037(5)$ for the series implementation, showing an increase in fidelity for the parallel implementation. 
With the parallel gate pulse duration of 300$\mu$s, as used in this result, we obtain roughly four-times speed up with the parallel implementation. 

\section{Comparison to prior methods}
\label{sec:TBD}

\begin{table*}[ht!]
\caption{\label{tab:Comparison} Comparison of various parallel pulse shaping methods.}
\begin{tabular}{| c| c| c| c| c| c| c|}
\hline
\multirow{2}{*} {Method} & {Pulse} & {Qubit} & {Calibration} & {Subset} & \multirow{2}{*}{Stability} & \multirow{2}{*} {Exactness} \\
{} & {Complexity} & {Overlap} & {Possibility} & {Graph Patterns} & {} &  \\
\hline
Ref.~\cite{figgatt2019parallel} & $o(GN)$ & No & Up to $O(N)$ & Yes & No & Not guaranteed \\
\hline
Ref.~\cite{grzesiak2020efficient} & $O(G+N)$ & Yes & Up to $O(N)$ & No & No${}^{*}$ & Yes\\
\hline
This Work & $O(G+N)$ & Yes & Yes & Yes & Yes & Yes\\
\hline
\hline
Series~\cite{blumel2021power} & $O(N)$ & N/A & Yes & N/A & Yes & Yes\\
\hline
\end{tabular}
\vspace{-1em}
\end{table*}

Table~\ref{tab:Comparison} summarizes a variety of metrics of interest for gate pulses, comparing our methods to those of prior methods~\cite{figgatt2019parallel,grzesiak2020efficient} that implement parallel entangling gates in trapped ions. For completeness, we further show applicable metrics for the series pulse in \cite{blumel2021power}. In the course of this work we investigated several related methods which are detailed in Appendices \ref{app:Method1}, \ref{app:Method2} and \ref{app:Method3}. Throughout this section, we use $G$ to denote the number of gates we implement in parallel. We first describe each metric before summarizing the advantages and disadvantages of each method with respect to these metrics.

{\it (Minimal) Pulse Complexity} -- We quantify the pulse complexity as the number of d.o.f. used to shape gate pulses, e.g., an individual pulse seen by each ion. In the amplitude modulated pulses~\cite{figgatt2019parallel,grzesiak2020efficient}, where the pulse function $g(t)$ is segmented with constant amplitudes with equal durations, the complexity corresponds to the number of pulse segments. In the amplitude and frequency modulated pulses such as the Fourier pulses in~\cite{blumel2021power} or ours, the complexity corresponds to the number of basis functions (e.g., Fourier components) used. 

{\it Qubit Overlap} -- This metric indicates whether a parallel-pulse method admits implementing an overlapping set of entangling gates, e.g., a gate over qubit pair (1,2), and a gate over (2,3), with the overlap at qubit 2.

{\it Calibration Possibility} -- In a fully connected system of $N$ ions, up to $\binom{N}{2}$ two-qubit entangling gates may be implemented in parallel. In a typical trapped-ion quantum computer calibration scheme, pulse amplitudes seen by individual ions are adjusted by some scaling factors to calibrate the entangling angle $\theta_{ij}$ for individual gates. The calibration possibility is then determined by how many gates may be calibrated according to the scheme above, for their simultaneous, parallel implementation. We use the notation ``Up to $O(N)$'' to indicate that there exists one or more instances where $O(N)$ gates may be calibrated, but not every gate patterns with $O(N)$ gates may be calibrated, for example a ring-graph pattern as discussed below and relevant to the data in Fig.~\ref{fig:CircuitsAndResultsHH}.

{\it Subset Graph Patterns} -- Once a set of pulses are determined for a parallel implementation of entangling gates, some methods may allow for a straightforward implementation of any subset of the entangling gates, while others do not. This possibility significantly reduces classical control overhead costs.

{\it Stability} -- It is desirable to stabilize gate pulses against drifts or fluctuations against experimental parameters~\cite{blumel2021power}. Here we indicate wether or not the stabilization strategies are compatible with the various methods considered. 

{\it Exactness} -- As discussed at length in Sec.~\ref{sec:Method} gate pulses must obey certain constraints to result in the desired gate unitaries. Exactness determines if the constraints are, at least on a theoretical level, met. 

As shown in Table~\ref{tab:Comparison} the method in~\cite{figgatt2019parallel} suffers from high pulse complexity. This results from the pulse shaping problem being cast as a non-linear optimization problem. The pulse solutions were obtained by using $o(GN)$ d.o.f. per problem construction, with a specific example demonstrated for two disjoint pairs of qubits. Extending the approach to more than two pairs in a manner consistent with the method of \cite{figgatt2019parallel}, i.e., recursively extending the method for each additional gate, the pulse complexity would result in $O(GN\exp(G))$. While this paper~\cite{figgatt2019parallel} did not explicitly consider the possibility of qubit overlap it is clear that it is not possible using a naive extension of the method. Consequently this limits the number of gates parallelizable to ``up to $O(N)$''. The $O(N)$ disjoint pairs of qubits may be straightforwardly calibrated using the amplitude scaling mentioned above. Regarding calibration possibility a subset of the disjoint pairs of qubits, assuming their parallel pulses are solved, can be implemented by simply omitting the pulses for the pairs in the complement set. The method did not consider stabilizing against experimental parameter changes. Finally, due to the non-linear problem formulation, the resulting pulse solution is not guaranteed to result in exactness.

In \cite{grzesiak2020efficient}, a linear solver is used to determine gate pulses, where $N$ d.o.f. are used to meet a set of constraints and $G$ additional d.o.f. were used to meet an additional set of constraints, necessitated by the goal to parallelize $G$ gates, leading to complexity $O(G+N)$. Qubit overlaps are allowed, increasing the number of entangling gates implementable in parallel to $\binom{N}{2}$ in principle, though the number of gates calibratable was still limited, up to $O(N)$ precluding certain graph patterns. Here we highlight an advantage of our method applied in the HH simulation example. Unlike the method in~\cite{grzesiak2020efficient} we can choose a global $J$ by tuning all $\theta_{ij}$ at will (see Appendix~\ref{app:CaliIssues} for details). In terms of subset graph patterns not every subset of the gates which were originally synthesized can be implemented straightforwardly because this approach solves gate pulses one qubit at a time not one gate at a time. For example, consider a set of pulses designed to entangle qubit pairs (1,2), (2,3), and (1,3), where the pulses were designed one qubit at a time; The pulses cannot trivially be modified (e.g., omit one pulse) to entangle (2,3), and (1,3) only. The pulses solved using the method of \cite{grzesiak2020efficient} were not originally designed with stability with respect to experimental parameter changes, although a latter method reported in \cite{blumel2021power} opened the door for the stability -- hence the asterisk in Table~\ref{tab:Comparison}. At last, the linear solver employed admits a theoretically exact pulse solution.

\section{Discussion}
\label{sec:Disc}
In this work we have demonstrated a flexible method to reduce total circuit execution time through parallelization of two-qubit entangling gate operations, highlighting their use in several well-known algorithms. In general practical benefits of gate parallelization materialize and scale as the number of gates that can be parallelized increases. At a relatively small number of gates parallelized, a cautionary note must be added, as a blindly-applied parallelization may in fact increase the circuit execution time~\cite{savage1998models}. Further, at present, the execution time is not the only metric to optimize, since a parallelized implementation may use more qubits, another precious resource. Additionally parallelization can induce more coherent errors than a serial implementation such as unintended entanglement. Our gate parallelization methodology would most benefit systems with short single qubit coherence times or high motional heating rates. A system with long motional dephasing times can further take advantage of our parallelization scheme by admitting lower power solutions (see Fig.~\ref{fig:PWRvsG}) or allowing for power rebalancing (see Sec.~\ref{sec:Method}).
In such a system parallelization would show even greater benefits from the reduction of the total circuit execution length. Note that the system as well as the algorithms used in Sec.~\ref{sec:Results} do not fully meet these criteria. Similarly, in terms of raw fidelity, the benefits of parallelization in any system dominated by coherent errors can be washed out however the total circuit execution time can still be reduced. 

In principle, our methodology provides a single pulse-solution set for all two-qubit entangling gates, without the need to synthesize new solutions for every gate combination. This unlocks critical memory efficiency: typical quantum control electronics include FPGA boards, used to store pulse shapes, known to have limited space. Prior methods~\cite{figgatt2019parallel,grzesiak2020efficient} would use prohibitively large memory for even a modest sized quantum computer, e.g., methods of~\cite{grzesiak2020efficient} would use $2^{O(N^2)}$ sets of pulses for different combinations. Our approach allows efficient calibration, where the pulses in the solution set are calibrated once and all subsets may be used without new calibration to implement corresponding parallel gates.

Currently there are two leading strategies for scaling the number of qubits in trapped-ion quantum computers: Increasing the number of ions in a trapping zone~\cite{chen2024benchmarking} or increasing the number of separate trapping zones~\cite{moses2023race}. The first method is challenged by long chains being difficult to control, due to for example, motional mode crowding and challenges in optical addressing. Meanwhile, disjoint, smaller ion chains suffer from limited parallelization capability due to the small number of qubits per trapping zone, in addition to significant shuttling times required to rearrange qubits that drastically hike quantum circuit execution time. As such, we believe there is a rich hardware optimization opportunity, where future state-of-the-art trapped-ion quantum computers may explore intermediate-sized per-chain number of qubits in a multi-chain architecture. 
Our parallel gate method, especially useful for a medium-length ion chain, would greatly enhance such an architecture.

Additionally, our method is compatible with other parallelization schemes that would further increase speed up. Since trapped-ion quantum computers may be configured to use multi-directional motional modes~\cite{zhu2023pairwise}, such as $X$, $Y$, and $Z$ directional modes, a total of $3P$ modes may be used simultaneously, further expanding the gate pulse design space.

Critical to high-fidelity implementations of our parallel pulses, as well as those implied by \cite{figgatt2019parallel,grzesiak2020efficient}, are accurate estimations of Lamb-Dicke parameters, as they are used as input to pulse solutions. Typical spectroscopic methods of estimating the parameters one at a time suffer from slow speed, often significantly eating into the up-time of a quantum computer~\cite{maksymov2022detecting}. However there are proposed protocols to make this process more efficient~\cite{kang2023efficient,liang2024pulse}. 

In this work we have shown the benefits of parallelization for near-term quantum computers however it will be even more important for fault tolerant quantum computers. Many syndrome based error correction schemes stand to benefit significantly from parallelization owing to the need for stabilizer readout and the nature of transversal logical gate operations~\cite{steane1998space,aharonov1997fault}. 

Further work to extend our approach may include finding ways to shorten the longer gate pulses that result from power rebalancing and power-time tradeoff when multiple gates overlap. For example, when $d$ gates form a star pattern, we obtain a ${\sim}\sqrt{d}$ speed up, shy of a full ${\sim} d$ speed up. In principle, it is possible to combine our parallelization technique with other gate optimizations such as reducing total power~\cite{blumel2021efficient}, and stabilizing $\chi_{ij}$ between multiple qubit pairs $(i,j)$~\cite{blumel2021power}. This would help provide the advantage of high-degree of stability exhibited in short ion chains to longer chains which exhibit a high-degree of parallelization. 

\section*{Acknowledgments}

N.M.L. acknowledges support from the Office of Naval Research under Awards N00014-23-1-2665, and N00014-24-1-2175. This material is based upon work supported by the U.S. Department of Energy, Office of Science, National Quantum Information Science Research Centers, Quantum Systems Accelerator (Award No. DE-SCL0000121). A.Y.N. acknowledges support from the National Science Foundation under Awards 2527952. Any opinion, finding, and conclusions or recommendations expressed in this material are those of the authors and do not necessarily reflect the views of ApexQuantum Inc. The authors would like to thank Anton Than for helpful comments. 

\bibliography{bib}

@article{debnath2016demonstration,
  title={Demonstration of a small programmable quantum computer with atomic qubits},
  author={Debnath, Shantanu and Linke, Norbert M and Figgatt, Caroline and Landsman, Kevin A and Wright, Kevin and Monroe, Christopher},
  journal={Nature},
  volume={536},
  number={7614},
  pages={63--66},
  year={2016},
  publisher={Nature Publishing Group UK London}
}

@article{childs2018toward,
  title={Toward the first quantum simulation with quantum speedup},
  author={Childs, Andrew M and Maslov, Dmitri and Nam, Yunseong and Ross, Neil J and Su, Yuan},
  journal={Proceedings of the National Academy of Sciences},
  volume={115},
  number={38},
  pages={9456--9461},
  year={2018},
  publisher={National Acad Sciences}
}

@article{grzesiak2020efficient,
  title={Efficient arbitrary simultaneously entangling gates on a trapped-ion quantum computer},
  author={Grzesiak, Nikodem and Bl{\"u}mel, Reinhold and Wright, Kenneth and Beck, Kristin M and Pisenti, Neal C and Li, Ming and Chaplin, Vandiver and Amini, Jason M and Debnath, Shantanu and Chen, Jwo-Sy and others},
  journal={Nature communications},
  volume={11},
  number={1},
  pages={2963},
  year={2020},
  publisher={Nature Publishing Group UK London}
}

@article{blumel2021power,
  title={Power-optimal, stabilized entangling gate between trapped-ion qubits},
  author={Bl{\"u}mel, Reinhold and Grzesiak, Nikodem and Pisenti, Neal and Wright, Kenneth and Nam, Yunseong},
  journal={npj Quantum Information},
  volume={7},
  number={1},
  pages={147},
  year={2021},
  publisher={Nature Publishing Group UK London}
}

@article{steane1998space,
  title={Space, time, parallelism and noise requirements for reliable quantum computing},
  author={Steane, Andrew M},
  journal={Fortschritte der Physik: Progress of Physics},
  volume={46},
  number={4-5},
  pages={443--457},
  year={1998},
  publisher={Wiley Online Library}
}

@book{savage1998models,
  title={Models of computation},
  author={Savage, John E},
  volume={136},
  year={1998},
  publisher={Addison-Wesley Reading, MA}
}

@inproceedings{aharonov1997fault,
  title={Fault-tolerant quantum computation with constant error},
  author={Aharonov, Dorit and Ben-Or, Michael},
  booktitle={Proceedings of the twenty-ninth annual ACM symposium on Theory of computing},
  pages={176--188},
  year={1997}
}

@article{blumel2021efficient,
  title={Efficient stabilized two-qubit gates on a trapped-ion quantum computer},
  author={Bl{\"u}mel, Reinhold and Grzesiak, Nikodem and Nguyen, Nhung H and Green, Alaina M and Li, Ming and Maksymov, Andrii and Linke, Norbert M and Nam, Yunseong},
  journal={Physical Review Letters},
  volume={126},
  number={22},
  pages={220503},
  year={2021},
  publisher={APS}
}

@article{bravyi2022constant,
  title={Constant-cost implementations of Clifford operations and multiply-controlled gates using global interactions},
  author={Bravyi, Sergey and Maslov, Dmitri and Nam, Yunseong},
  journal={Physical Review Letters},
  volume={129},
  number={23},
  pages={230501},
  year={2022},
  publisher={APS}
}

@article{grzesiak2022efficient,
  title={Efficient quantum programming using EASE gates on a trapped-ion quantum computer},
  author={Grzesiak, Nikodem and Maksymov, Andrii and Niroula, Pradeep and Nam, Yunseong},
  journal={Quantum},
  volume={6},
  pages={634},
  year={2022},
  publisher={Verein zur F{\"o}rderung des Open Access Publizierens in den Quantenwissenschaften}
}

@inproceedings{maksymov2022detecting,
  title={Detecting Qubit-coupling faults in ion-trap quantum computers},
  author={Maksymov, Andrii and Nguyen, Jason and Chaplin, Vandiver and Nam, Yunseong and Markov, Igor L},
  booktitle={2022 IEEE International Symposium on High-Performance Computer Architecture (HPCA)},
  pages={387--399},
  year={2022},
  organization={IEEE}
}

@article{figgatt2019parallel,
  title={Parallel entangling operations on a universal ion-trap quantum computer},
  author={Figgatt, Caroline and Ostrander, Aaron and Linke, Norbert M and Landsman, Kevin A and Zhu, Daiwei and Maslov, Dmitri and Monroe, Christopher},
  journal={Nature},
  volume={572},
  number={7769},
  pages={368--372},
  year={2019},
  publisher={Nature Publishing Group UK London}
}

@article{BV_Algo,
author = {Bernstein, Ethan and Vazirani, Umesh},
title = {Quantum Complexity Theory},
journal = {SIAM Journal on Computing},
volume = {26},
number = {5},
pages = {1411-1473},
year = {1997},
doi = {10.1137/S0097539796300921},
}

@article{Par_Fid_1,
  title={Experimental entanglement of four particles},
  author={Sackett, Cass A and Kielpinski, David and King, Brian E and Langer, Christopher and Meyer, Volker and Myatt, Christopher J and Rowe, M and Turchette, QA and Itano, Wayne M and Wineland, David J and others},
  journal={Nature},
  volume={404},
  number={6775},
  pages={256--259},
  year={2000},
  publisher={Nature Publishing Group UK London}
}

@article{Par_Fid_2,
  title = {High-Fidelity Quantum Logic Gates Using Trapped-Ion Hyperfine Qubits},
  author = {Ballance, C. J. and Harty, T. P. and Linke, N. M. and Sepiol, M. A. and Lucas, D. M.},
  journal = {Phys. Rev. Lett.},
  volume = {117},
  issue = {6},
  pages = {060504},
  numpages = {6},
  year = {2016},
  month = {Aug},
  publisher = {American Physical Society},
  doi = {10.1103/PhysRevLett.117.060504},
  url = {https://link.aps.org/doi/10.1103/PhysRevLett.117.060504}
}

@article{kang2023efficient,
  title={Efficient motional-mode characterization for high-fidelity trapped-ion quantum computing},
  author={Kang, Mingyu and Liang, Qiyao and Li, Ming and Nam, Yunseong},
  journal={Quantum Science and Technology},
  volume={8},
  number={2},
  pages={024002},
  year={2023},
  publisher={IOP Publishing}
}

@article{van2006quantum,
  title={Quantum algorithms for some hidden shift problems},
  author={Van Dam, Wim and Hallgren, Sean and Ip, Lawrence},
  journal={SIAM Journal on Computing},
  volume={36},
  number={3},
  pages={763--778},
  year={2006},
  publisher={SIAM}
}

@inproceedings{rotteler2010quantum,
  title={Quantum algorithms for highly non-linear Boolean functions},
  author={R{\"o}tteler, Martin},
  booktitle={Proceedings of the twenty-first annual ACM-SIAM symposium on Discrete algorithms},
  pages={448--457},
  year={2010},
  organization={SIAM}
}

@article{lu2019global,
  title={Global entangling gates on arbitrary ion qubits},
  author={Lu, Yao and Zhang, Shuaining and Zhang, Kuan and Chen, Wentao and Shen, Yangchao and Zhang, Jialiang and Zhang, Jing-Ning and Kim, Kihwan},
  journal={Nature},
  volume={572},
  number={7769},
  pages={363--367},
  year={2019},
  publisher={Nature Publishing Group UK London}
}

@article{evered2023high,
  title={High-fidelity parallel entangling gates on a neutral-atom quantum computer},
  author={Evered, Simon J and Bluvstein, Dolev and Kalinowski, Marcin and Ebadi, Sepehr and Manovitz, Tom and Zhou, Hengyun and Li, Sophie H and Geim, Alexandra A and Wang, Tout T and Maskara, Nishad and others},
  journal={Nature},
  volume={622},
  number={7982},
  pages={268--272},
  year={2023},
  publisher={Nature Publishing Group UK London}
}

@article{levine2019parallel,
  title={Parallel implementation of high-fidelity multiqubit gates with neutral atoms},
  author={Levine, Harry and Keesling, Alexander and Semeghini, Giulia and Omran, Ahmed and Wang, Tout T and Ebadi, Sepehr and Bernien, Hannes and Greiner, Markus and Vuleti{\'c}, Vladan and Pichler, Hannes and others},
  journal={Physical review letters},
  volume={123},
  number={17},
  pages={170503},
  year={2019},
  publisher={APS}
}

@article{arute2019quantum,
  title={Quantum supremacy using a programmable superconducting processor},
  author={Arute, Frank and Arya, Kunal and Babbush, Ryan and Bacon, Dave and Bardin, Joseph C and Barends, Rami and Biswas, Rupak and Boixo, Sergio and Brandao, Fernando GSL and Buell, David A and others},
  journal={Nature},
  volume={574},
  number={7779},
  pages={505--510},
  year={2019},
  publisher={Nature Publishing Group UK London}
}

@article{Adder,
author = {Draper, Thomas G. and Kutin, Samuel A. and Rains, Eric M. and Svore, Krysta M.},
title = {A logarithmic-depth quantum carry-lookahead adder},
year = {2006},
issue_date = {July 2006},
publisher = {Rinton Press, Incorporated},
address = {Paramus, NJ},
volume = {6},
number = {4},
issn = {1533-7146},
journal = {Quantum Info. Comput.},
month = jul,
pages = {351–369},
numpages = {19}
}

@article{sorensen1999quantum,
  title={Quantum computation with ions in thermal motion},
  author={S{\o}rensen, Anders and M{\o}lmer, Klaus},
  journal={Physical review letters},
  volume={82},
  number={9},
  pages={1971},
  year={1999},
  publisher={APS}
}

@article{molmer1999multiparticle,
  title={Multiparticle entanglement of hot trapped ions},
  author={M{\o}lmer, Klaus and S{\o}rensen, Anders},
  journal={Physical Review Letters},
  volume={82},
  number={9},
  pages={1835},
  year={1999},
  publisher={APS}
}

@article{sorensen2000entanglement,
  title={Entanglement and quantum computation with ions in thermal motion},
  author={S{\o}rensen, Anders and M{\o}lmer, Klaus},
  journal={Physical Review A},
  volume={62},
  number={2},
  pages={022311},
  year={2000},
  publisher={APS}
}

@article{bonechi1992heisenberg,
  title={Heisenberg XXZ model and quantum Galilei group},
  author={Bonechi, F and Celeghini, E and Giachetti, Riccardo and Sorace, E and Tarlini, M},
  journal={Journal of Physics A: Mathematical and General},
  volume={25},
  number={15},
  pages={L939},
  year={1992},
  publisher={IOP Publishing}
}

@article{maslov2017basic,
  title={Basic circuit compilation techniques for an ion-trap quantum machine},
  author={Maslov, Dmitri},
  journal={New Journal of Physics},
  volume={19},
  number={2},
  pages={023035},
  year={2017},
  publisher={IOP Publishing}
}

@article{suzuki1991general,
  title={General theory of fractal path integrals with applications to many-body theories and statistical physics},
  author={Suzuki, M.},
  journal={Journal of Mathematical Physics},
  volume={32},
  number={2},
  pages={400--407},
  year={1991},
  doi={10.1063/1.529425},
  url={https://aip.scitation.org/doi/10.1063/1.529425},
  publisher={American Institute of Physics}
}

@article{liang2024pulse,
  title={Pulse optimization for high-precision motional-mode characterization in trapped-ion quantum computers},
  author={Liang, Qiyao and Kang, Mingyu and Li, Ming and Nam, Yunseong},
  journal={Quantum Science and Technology},
  volume={9},
  number={3},
  pages={035007},
  year={2024},
  publisher={IOP Publishing}
}

@article{zhu2023pairwise,
  title={Pairwise-Parallel Entangling Gates on Orthogonal Modes in a Trapped-Ion Chain},
  author={Zhu, Yingyue and Green, Alaina M and Nguyen, Nhung H and Huerta Alderete, C and Mossman, Elijah and Linke, Norbert M},
  journal={Advanced Quantum Technologies},
  volume={6},
  number={11},
  pages={2300056},
  year={2023},
  publisher={Wiley Online Library}
}

@article{duan2014linear,
  title={Linear-time approximation for maximum weight matching},
  author={Duan, Ran and Pettie, Seth},
  journal={Journal of the ACM (JACM)},
  volume={61},
  number={1},
  pages={1--23},
  year={2014},
  publisher={ACM New York, NY, USA}
}

@article{chen2024benchmarking,
  title={Benchmarking a trapped-ion quantum computer with 30 qubits},
  author={Chen, Jwo-Sy and Nielsen, Erik and Ebert, Matthew and Inlek, Volkan and Wright, Kenneth and Chaplin, Vandiver and Maksymov, Andrii and P{\'a}ez, Eduardo and Poudel, Amrit and Maunz, Peter and others},
  journal={Quantum},
  volume={8},
  pages={1516},
  year={2024},
  publisher={Verein zur F{\"o}rderung des Open Access Publizierens in den Quantenwissenschaften}
}

@article{moses2023race,
  title={A race-track trapped-ion quantum processor},
  author={Moses, Steven A and Baldwin, Charles H and Allman, Michael S and Ancona, R and Ascarrunz, L and Barnes, C and Bartolotta, J and Bjork, B and Blanchard, P and Bohn, M and others},
  journal={Physical Review X},
  volume={13},
  number={4},
  pages={041052},
  year={2023},
  publisher={APS}
}

@misc{WebsiteToFiles,
title = {Mode and Eta Values},
note = {\url   {https://github.com/athan42-umd/mode-parameters}}
}

\appendix

\section{Different parallel pulse protocols}
\label{app:ParaMethods}

In this section, we show three different protocols that may be used to implement parallel two-qubit gates in a trapped-ion quantum computer. In particular, we discuss their implementation details and pros and cons. We note the protocol used in the main text appears in Sec.~\ref{app:Method3}.

\subsection{Sequencing protocol}
\label{app:Method1}
Consider a scenario where two-qubit gate pulses $g(t)$ are solved individually such that each gate pulse, when implemented, results in a single two-qubit gate. Denoting the different pulses for different pairs as $g_\nu(t)$ with index $\nu \in \{1,2,..,{\mathcal N}\}$, the following steps may be used to implement ${\mathcal N}$ two-qubit gates simultaneously. Hereafter, we assume ${\mathcal N}$ is a power of two and the target two-qubit gates always induce fully entangling phases, i.e., $\chi = \pi/2$ for every qubit pair to be entangled, for simplicity.

\begin{enumerate}
\item Let $g_\nu(t)$ be a gate pulse that, when implemented, results in $\chi_\nu = (\pi/2)$ for every $\nu$. For example, pulses $g_\nu(t)$ may be solved using Fourier basis functions, as in \cite{blumel2021efficient,blumel2021power}, admitting power optimality.
\item Synthesize a pulse sequence $G_{\mathcal N} := (s_1,s_2,..,s_{\mathcal N})$ for ${\mathcal N}$ pairs, where $s_m$ are the sequences for pair $m$, with the following recursion relation: $G_{\mathcal N} = (G_{{\mathcal N}/2}, G_{{\mathcal N}/2}) \mathbin\Vert (G_{{\mathcal N}/2}, -G_{{\mathcal N}/2})$, where $\mathbin\Vert$ denotes concatenation and it is assumed that, in an example sequence $(G_{{\mathcal N}/2}, -G_{{\mathcal N}/2})$, $G_{{\mathcal N}/2}$ is applicable to the first ${\mathcal N}/2$ pairs and $-G_{{\mathcal N}/2}$ is applicable to the last ${\mathcal N}/2$ pairs. $G_1 = +1$.
\item Apply $g_\nu(t)$ for each $\nu$ with signs implied by $G_{\mathcal N}$. For each pair $\nu$, the sequence is of length ${\mathcal N}$.
\end{enumerate}

The above sequence-based protocol hinges on the fact that, when applying $-g_\nu(t)$ to each qubit of pair $\nu$, the induced entanglement phase angle is still $\chi_\nu$, not $-\chi_\nu$, and a negative phase angle is obtained only when two qubits see opposite-signed pulses. The sequences change the signs in such a way that all undesirable couplings have entanglement phases that are canceled out. The overall sequence uses ${\mathcal N}$ gate pulses per pair, with each gate pulse inducing $(\pi/2)/{\mathcal N}$ phases. For a 1-regular graph pattern of two-qubit gates to implement simultaneously, with power-time trade off mentioned in the main text, each gate pulse may be made 1/$\sqrt{{\mathcal N}}$ times the duration of a single gate pulse with a full entanglement phase angle, leading to a speed up of ${\sim} \sqrt{{\mathcal N}}$ for implementing a total of ${\mathcal N}$ two-qubit gates using the sequencing protocol. For a star graph pattern, the speed up is negligible, since the nominal ${\mathcal N}$ speed up is countered by ${\mathcal N}$ slow down due to power rebalancing detailed in the main text.

Since the gate pulses synthesized in this section may use the series pulses ${\mathcal N}$ times, the overall number of d.o.f. used becomes $O({\mathcal N}n)$, as each series pulse uses $O(n)$ d.o.f. as per \cite{blumel2021power}, where $n$ is the number of ions in an ion chain. Arbitrary patterns of gates may be implemented in parallel, allowing for qubit overlaps and each gate may be calibrated individually as a series gate. Subsets of the gates whose pulses are synthesized for may be implemented straightforwardly by simply omitting the pulses corresponding to the complement sets. Since the series gates of \cite{blumel2021efficient,blumel2021power} admit stabilization against experimental parameter drifts or fluctuations (see Sec.~\ref{App:Stabilization}), the gate pulses herein also admit the stability. Similarly, the theoretical exactness of the pulses afforded by \cite{blumel2021efficient,blumel2021power} translates directly to the pulses herein.

\subsection{Disjoint null-space protocol}
\label{app:Method2}
Another method to implement ${\mathcal N}$ two-qubit entangling gates simultaneously may be as follows. Recall from main text our basis function of choice is sine functions and that the pulse constraints include $\alpha_{ip}=0$. We may opt to satisfy these constraints by first assigning sine functions with differerent frequencies $l/\tau$ into ${\mathcal N}$ mutually exclusive sets, then solving for $\alpha_{ip}=0$ for each set by finding the null space in a similar manner as in the main text. In each of the ${\mathcal N}$ null space, a power-optimal pulse for an entangling gate may be found, again in a similar manner as in the main text. Which sine functions to assign to which set and which set to use to couple which pair of qubits may be chosen heuristically, for example based on the size of Lamb-Dicke parameters associated with a pair of qubits to entangle for a main mode of choice to use to induce the entanglement. 

To see why the above approach works, we start by two pulse solutions $g(t)$ and $h(t)$, where $g(t)$ is a linear superposition of sine function elements of a set and $h(t)$ is a linear superposition of sine function elements of a different set. Such pulses admit
\begin{equation}
\label{eq:OrthPulse}
\int_0^\tau g(t) h(t) dt = 0.
\end{equation}
Further, by construction, the pulses further admit
\begin{equation}
\label{eq:PhaseClose}
\int_0^\tau g(t) e^{i\omega_p t} dt = 0, \quad
\int_0^\tau h(t) e^{i\omega_p t} dt = 0,
\end{equation}
which are equivalent to $\alpha_{ip}=0$ conditions.
For these pulse solutions to work, i.e., when the two pulses are applied to two different pairs of qubits, the resulting unitary applied to a quantum computer is a parallel implementation of two entangling gates for the two pairs, we need to ensure that $\chi_{ij}=0$, where qubit $i$ is drawn from one pair (say, $g(t)$ is applied) and qubit $j$ is drawn from the other pair (say, $h(t)$ is applied). One way to achieve this is to satisfy $\theta_{ijp} = 0$ [see main text equation (\ref{eq:timeevolution})], which may be written as
\begin{equation}
\label{eq:DisjointCrosstalk}
\theta_{ijp} = 4\int_0^\tau dt_1 g(t_1) \int_0^\tau dt_2 h(t_2) \sin(\omega_p (t_2-t_1)) = 0.
\end{equation}
Expanding $\sin(\omega_p (t_2-t_1))$ in (\ref{eq:DisjointCrosstalk}) with the well-known angle subtraction formula and inserting 
\begin{equation} 
\label{eq:ghExp}
g(t_1) = \sum_k a_k \sin\left(\frac{2\pi k}{\tau} t_1\right), \;
h(t_2) = \sum_l b_l \sin\left(\frac{2\pi l}{\tau} t_2\right)
\end{equation}
in (\ref{eq:DisjointCrosstalk}), we obtain
\begin{align}
\label{eq:theta_ijp_intermediate}
\theta_{ijp} = &4\sum_{k,l} a_k b_l \int_0^\tau dt_1 \sin\left(\frac{2\pi k}{\tau} t_1\right) \frac{1}{\left(\frac{2\pi l}{\tau}\right)^2-\omega_p^2} \times \nonumber \\
\Bigg[&\cos(\omega_p t_1)\cos(\omega_p t_1) \left( \omega_p \sin\left( \frac{2\pi l}{\tau}t_1\right)\right) \nonumber  \\
-&\cos(\omega_p t_1)\sin(\omega_p t_1) \left( \frac{2\pi l}{\tau} \cos\left( \frac{2\pi l}{\tau}t_1\right) \right) \nonumber  \\
+&\sin(\omega_p t_1)\cos(\omega_p t_1) \left( \frac{2\pi l}{\tau} \cos\left( \frac{2\pi l}{\tau}t_1\right) \right)  \nonumber \\
+&\sin(\omega_p t_1)\sin(\omega_p t_1) \left( \omega_p \sin\left( \frac{2\pi l}{\tau}t_1\right)\right)  \nonumber  \\
-&\sin(\omega_p t_1) \frac{2\pi l}{\tau} \Bigg],
\end{align}
where we used
\begin{align}
&\int_0^{t_1} dt_2 \sin\left(\frac{2\pi l}{\tau} t_2\right) \sin(\omega_p t_2) = \nonumber \\ & 
\frac{\omega_p \cos(\omega_p t_1)\sin\left(\frac{2\pi l}{\tau} t_1\right)-\frac{2\pi l}{\tau}\cos\left(\frac{2\pi l}{\tau} t_1\right)\sin(\omega_p t_1)}{\left(\frac{2\pi l}{\tau}\right)^2 - \omega_p^2}
\end{align}
and
\begin{align}
&\int_0^{t_1} dt_2 \sin\left(\frac{2\pi l}{\tau} t_2\right) \cos(\omega_p t_2) = \nonumber \\ & 
\frac{\frac{2\pi l}{\tau}-\frac{2\pi l}{\tau}\cos\left(\frac{2\pi l}{\tau} t_1\right)\cos(\omega_p t_1)-\omega_p \sin(\omega_p t_1)\sin\left(\frac{2\pi l}{\tau} t_1\right)}{\left(\frac{2\pi l}{\tau}\right)^2 - \omega_p^2}.
\end{align}
$\theta_{ijp}$ in (\ref{eq:theta_ijp_intermediate}) may be simplified to
\begin{align}
\theta_{ijp} =& 4\sum_{k,l} \frac{a_k b_l}{\left(\frac{2\pi l}{\tau}\right)^2-\omega_p^2} \times \nonumber \\ \Bigg[&\omega_p \int_0^\tau dt_1 \sin\left(\frac{2\pi k}{\tau} t_1 \right) \sin\left( \frac{2\pi l}{\tau} t_1 \right) \nonumber \\ 
-&\frac{2\pi l}{\tau} \int_0^\tau dt_1 \sin\left( \frac{2\pi k}{\tau} t_1 \right) \sin(\omega_p t_1) \Bigg] \nonumber \\
=& 0,
\end{align}
where the first integral evaluates to zero since $k \neq l$ by construction [see also (\ref{eq:OrthPulse}) and (\ref{eq:ghExp})] and the second integral evaluates to zero as well [see (\ref{eq:PhaseClose})].

While the method of first separating the basis functions into different frequency bands, then solving for the null space to satisfy $\alpha_{ip} = 0$, results in the pulses that satisfy all the constraints, the pulses may exhibit unfavorable features, such as a large power requirement or long gate duration. For example, a single entangling gate admits a reasonable level of power when the gate duration is larger than $\Delta \omega_p^{-1}$~\cite{blumel2021efficient}, where $\Delta \omega_p$ is a typical mode frequency difference between two neighboring motional modes. This may be understood as having a sufficient number of degrees of freedom or frequency components near the ion-chain motional mode frequencies, considering there are $\alpha_{ip}=0$ constraints to satisfy, of which there are $n$. The gate protocol used in this section effectively does the same, except that the constraints are satisfied for ${\mathcal N}$ times using mutually exclusive frequency bands for each gates. This results in the minimal gate duration that scales as $O({\mathcal N}n)$, for admitting a reasonable power requirement. If the gate duration is to be comfortably above $\Delta \omega_p^{-1}$ for any reasons, including hardware limitations, the protocol of this section may in fact be employed to result in useful pulses for a modest sized ${\mathcal N}$, commensurate with how much longer a typical gate duration is when compared to $\Delta \omega_p^{-1}$. In a quantum computer engineered to deliver computational results as fast as possible though, the protocol may not be as useful.

To recap, based on the discussion in the above paragraph, the overall number of d.o.f. used becomes $O({\mathcal N}n)$. Arbitrary patterns of gates may be implemented in parallel, allowing for qubit overlaps and each gate may be calibrated individually since crosstalks induced between any two pulses is naturally null by design. As in the previous protocol, subsets of the gates whose pulses are synthesized for may be implemented straightforwardly by simply omitting the pulses corresponding to the complement sets. Since each of the constituent gate may admit stabilization against experimental parameter drifts or fluctuations following the same approach as in Sec.~\ref{App:Stabilization}, the gate pulses herein also admit the stability. The theoretical exactness of the pulses results from the pulse design method detailed in this section.

\subsection{Common null-space protocol}
\label{app:Method3}
The protocol used in the main text is the common null-space protocol, detailed in this section. Unlike the disjoint null-space protocol detailed in the previous section, here, the null space that satisfies $\alpha_{ip}=0$ is solved once, over the entire frequency range of the sine basis functions, the same as a conventional single entangling gate pulse synthesis method. Specifically, denoting the linear combinations of the basis functions that obey $\alpha_{ip}=0$ conditions as new basis functions $\vec{a}$, i.e., $\hat{M}\vec{a} = \vec{0}$, where $\hat{M}$ is the matrix that encodes $\alpha_{ip}=0$ conditions as defined in the main text Sec.~\ref{sec:Method}, the gate pulses may be drawn from the space spanned by the (orthonormalized) null space vectors $\vec{a}$ of $\hat{M}$. Any gate pulses drawn from such a space automatically satisfies $\alpha_{ip}=0$, i.e., the pulses ideally perfectly decouple qubits from the motional modes at the end of the gate.

Denoting now the row vector of these null space vectors $\vec{a}$ as $\hat{A}$, $\theta_{ijp}$ in the main text Eq.~(\ref{eq:timeevolution}) may be computed as $\vec{r}_i^T \hat{R}_p \vec{r}_j$, where $\vec{r}_i$ and $\vec{r}_j$ denote the gate pulses for qubits $i$ and $j$, respectively, in the null space basis and $\hat{R}_p {=} \hat{A}^T \hat{S}_p \hat{A}$ is the matrix defined in the main text Sec.~\ref{sec:Method}; To obtain a sine basis gate pulse $\vec{v}_i$ from the null space basis gate pulse $\vec{r}_i$, we may compute $\vec{v}_i = \hat{A}\vec{r}_i$. Provided with the method to compute $\theta_{ijp}$, the remaining steps of the protocol largely consist of two: ``preprocessing'' to determine which mode $p$ to (mainly) use to couple qubits $i$ and $j$ based on $\theta_{ijp}$, which approximately solves for the gate pulses most of the way, then finalizing the gate pulses based on the preprocessing results, which solves for the gate pulses precisely.
More concretely, we first compute eigenvalue-eigenvector pairs for $R_p$, ordering them in the descending order of the modulus of the eigenvalues. Focus for example on the largest modulus eigenvalue and the associated eigenvector. The eigenvector, if used as the gate pulse for qubits $i$ and $j$, induces entanglement between $i$ and $j$, where most of its amount comes from $\theta_{ijp}$, whose value will be the largest-modulus eigenvalue. The large eigenvalue translates to efficient power requirement for gate pulses and the first several largest modulus eigenvalues may be gleaned from the ordered list of eigenvalues mentioned above. As such, we may tabulate $\eta_{ip}\eta_{jp}\theta_{ijp}$ values for various $\lambda$ and find pairings between couplings $(i,j)$ and mode-eigenvalue $(p,\lambda)$ such that $\eta_{ip}\eta_{jp}\theta_{ijp}$ for the chosen pairings are large throughout; note we have suppressed the dependency of $\theta_{ijp}$ on $\lambda$ for notational brevity. This effectively achieves low pulse power consumed for coupling $i$ and $j$, via mode $p$, when the gate pulses are rescaled to roughly obtain the target entanglement angles, in the regime where gate duration $\tau$ is larger than the characteristic time scale implied by the motional mode frequency spacing -- in this regime, a gate pulse that strongly interacts with one mode does not strongly interact with another mode. The inexactness in the angles obtained compared with the target angles arise from the small angles that arise from coupling qubits $i$ and $j$ via modes $p'\neq p$, i.e., $\chi_{ij} = \theta_{ijp} +  \sum_{p'\neq p} \theta_{ijp'}$.

Once the parings between $(i,j)$ and $p$ are made, for example by casting the pairing problem described above to a maximum weight matching problem maximizing the sum of $\eta_{ip}\eta_{jp}\theta_{ijp}$, the pairings determined may be used for the final step of solving for the gate pulses exactly. In particular, we start with the pairing that admits the smallest modulus $\eta_{ip}\eta_{jp}\theta_{ijp}$ value, and assign the corresponding eigenvector $\vec{r}_{ij}$ to be the fundamental shape of the gate pulse. The entanglement angle obtained for qubits $i$ and $j$ will be $\chi_{ij} = \vec{r}_{ij}^T (\sum_p \eta_{ip}\eta_{jp} \hat{R}_p) \vec{r}_{ij}$. We then adjust the norm of $\vec{r}_{ij}$ such that the rescaled $\chi_{ij}$ value matches the target $\chi_{ij}$ value. For the second pairing, say $(k,l)$, and $(i,j)$ and $(k,l)$ are assumed to be disjoint for example-description purposes, we may not in general use the eigenvectors as are for our gate pulses $\vec{r}_{kl}$. This is so, since there may be a non-zero contribution to $\chi_{ik}$, $\chi_{il}$, $\chi_{jk}$, and $\chi_{jl}$ that stem from pulses $\vec{r}_{ij}$ and $\vec{r}_{kl}$. To solve for appropriate $\vec{r}_{kl}$, we constrain the pulse design space of $\vec{r}_{kl}$ such that any pulse drawn from the space automatically satisfies crosstalk terms $\chi_{ik}$, $\chi_{il}$, $\chi_{jk}$, and $\chi_{jl}$ to be zero. To do so, we modify $\hat{R}_p$ used above to $\hat{R}^*_p$, where $\hat{R}^*_p = \hat{Q}^T \hat{S}_p \hat{Q}$ and $\hat{Q}$ is the projector, to be defined, that achieves the design-space constraint. We compute the projector $\hat{Q}$ to be $(1-\sum_{q} \vec{w}_{q} \vec{w}_{q})^T$, where $\vec{w}_{q}$, $q\in [0,4)$ are the orthonormalized vectors of $\sum_p \eta_{i,p}\eta_{k,p} \hat{R}_p \vec{r}_{ij}$, $\sum_p \eta_{i,p}\eta_{l,p} \hat{R}_p \vec{r}_{ij}$, $\sum_p \eta_{j,p}\eta_{k,p} \hat{R}_p \vec{r}_{ij}$, and $\sum_p \eta_{j,p}\eta_{l,p} \hat{R}_p \vec{r}_{ij}$. More generally, in case there is an overlap between $(i,j)$ and $(k,l)$, we may skip the overlapping qubit cases, i.e., if $i=k$, we may skip a vector $\sum_p \eta_{i,p}\eta_{k,p} \hat{R}_p \vec{r}_{ij}$ from projector $\hat{Q}$. We may iterate this process for subsequent couplings with an expanding set of crosstalk terms to consider.

The gate pulses synthesized according to the protocol detailed in this section uses $O(n)$ d.o.f. to satisfy $\alpha_{ip}=0$ conditions and additional $O({\mathcal N})$ d.o.f. to satisfy $\theta_{jp}$ conditions, where $n$ is the number of ions in an ion chain and ${\mathcal N}$ is the number of two-qubit entangling gates parallelized. Arbitrary patterns of gates may be implemented in parallel and individual pulses may be calibrated independently owing to the nulling of the crosstalk between any two pulses obtained via projectors $\hat{Q}$. Consequently, any subset of the gates whose pulses are synthesized for may be implemented straightforwardly by, once again, omitting the gate pulses belonging to the complement set. Stabilization and exactness follows from the similar argument as the two of our other methods before.

\section{Properties of the parallel pulses}

\subsection{Power Scaling}
\label{app:PowerScaling}

\begin{figure*}
\centering
(a)\includegraphics[width=0.45\textwidth]{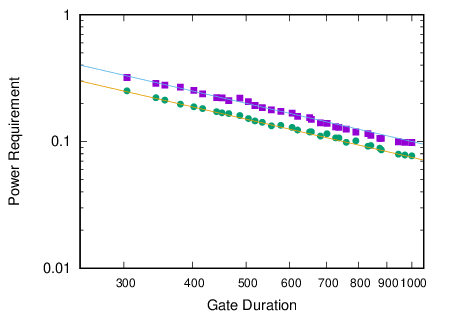} 
(b)\includegraphics[width=0.45\textwidth]{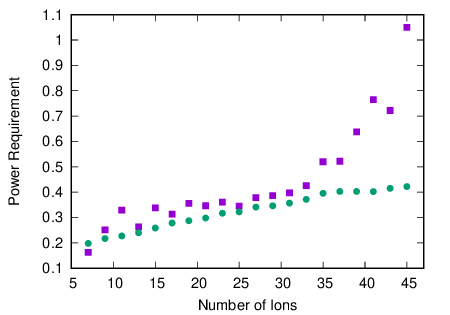} 
    \caption{\label{fig:Power}
    Power scaling $\bar{g}$ of parallel vs. serial gates. Purple squares represent parallel, green circles represent serial. (a) Simulations with $N=13$. For parallel, we consider $N$ sets of $\frac{N-1}{2}$ gates, each set being parallelized. For serial, all pairs are considered. The largest power solution for parallel and serial is plotted. One-to-one power-time tradeoff is observed for both implementations. (b) Power requirements as a function of the number of ions. Pulses were a fixed duration synthesized with $\tau=300\mu s$. Only the largest power requirements are reported; for serial implementation we consider all $\frac{N(N-1)}{2}$ gates. For parallel, we report the largest value out of randomly selected $\frac{(N-1)}{2}$ couplings being parallelized.
}
\end{figure*}

\begin{figure}[h]
\includegraphics{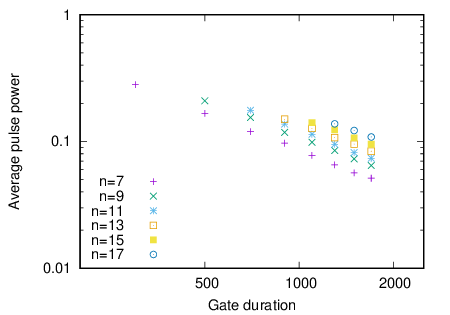}
\caption{\label{fig:PWRTIME} Average pulse power $\langle \bar{g} \rangle$ as a function of gate duration $\tau$ for various number of ions. All $\binom{n}{2}$ gates are solved for their simultaneous implementation and the pulse power requirement $\bar{g}$ is averaged over all gates.}
\end{figure}

Figure~\ref{fig:Power}(a) shows pulse power requirement as a function of gate duration for a chosen number of ions $N=13$. We use an ideal set of mode frequencies and Lamb-Dicke parameters obtained by assuming  We observe that there is a one-to-one power-time tradeoff, just as in the series pulses. This allows us to implement the power rebalancing technique without incurring too much of a power requirement on any one of the ions at the expense of longer gate duration. See also Fig.~\ref{fig:PWRTIME} for the all-to-all coupling cases. In order to next investigate the power scaling of our pulses for a fixed duration pulse, we show in Fig.~\ref{fig:Power}(b) the pulse power requirement, measured as the largest power required across all gate pulses considered, as a function of the number of ions; Considered therein are the series gates for all pairs and the parallel gates for an instance of $\lfloor n/2 \rfloor$ disjoint gates. We observe the power scaling of $\sim N$ up until $N=35$, and a more rapid increase thereafter for pulses of $\tau=300 \mu \rm{s}$. This is consistent with our understanding, since for $N\ge35$ the mode-frequency spacing becomes smaller than $3 \rm{kHz}$, the frequency scale commensurate with $\tau=300\mu \rm{s}$, crowding out some gates from strong couplings to motional modes.

Tables~\ref{tab:mode13} and~\ref{tab:eta13} show the mode frequencies and Lamb-Dicke parameters for various number of ions. For $n > 13$, we refer the readers to a repository made available in \cite{WebsiteToFiles}.

\subsection{Motional-Mode Stabilization}
\label{App:Stabilization}
Our approach allows for additional constraints to be added to the gate pulse solver as in previous methods~\cite{blumel2021power}. One such example is to stabilize the gate pulses, in particular ion-mode decoupling represented by $\alpha_{ip}$, against motional-mode frequency drifts. This is achieved by modifying the $\alpha_{ip}$ constraint in~(\ref{eq:timeevolution}) to be 
\begin{align}
\label{eq:Stabilization}
    \frac{\partial^k\alpha_{ip}}{\partial\omega_p^k} =
    \frac{\partial^k}{\partial\omega ^k_p}\int_0^\tau g(t)e^{i\omega_pt}dt +h.c.=0,
\end{align}
where $k=1,2,...,K$, and $K$ denotes the desired stabilization order. Since the constraints in (\ref{eq:Stabilization}) are linear, all we have to do to stabilize the pulse up to the $K^{th}$ order is to just add these additional constraint equations into the coefficient matrix $M_{pl}$ used in the main text.

\begin{table*}
\caption{Lamb-Dicke parameters for seven ions.}
\label{tab:eta7}
\scriptsize
\centering
\begin{tabular}{lrrrrrrr}
\toprule
 & Mode 1 & Mode 2 & Mode 3 & Mode 4 & Mode 5 & Mode 6 & Mode 7 \\
\midrule
Ion 1 & 0.01079 & 0.02317 & -0.03623 & 0.04795 & 0.05655 & -0.05987 & 0.04233  \\
Ion 2 & -0.03503 & -0.05859 & 0.05854 & -0.03174 & 0.01021 & -0.04597 & 0.04233  \\
Ion 3 & 0.05720 & 0.05243 & 0.00789 & -0.05666 & -0.03780 & -0.02491 & 0.04233  \\
Ion 4 & -0.06593 & 0.00000 & -0.06038 & 0.00000 & -0.05793 & -0.00000 & 0.04233  \\
Ion 5 & 0.05720 & -0.05243 & 0.00789 & 0.05666 & -0.03780 & 0.02491 & 0.04233  \\
Ion 6 & -0.03503 & 0.05859 & 0.05854 & 0.03174 & 0.01021 & 0.04597 & 0.04233  \\
Ion 7 & 0.01079 & -0.02317 & -0.03623 & -0.04795 & 0.05655 & 0.05987 & 0.04233  \\
\bottomrule
\end{tabular}
\end{table*}

\begin{table*}
\caption{Lamb-Dicke parameters for nine ions.}
\label{tab:eta9}
\scriptsize 
\centering
\begin{tabular}{lrrrrrrrrr}
\toprule
 & Mode 1 & Mode 2 & Mode 3 & Mode 4 & Mode 5 & Mode 6 & Mode 7 & Mode 8 & Mode 9 \\
\midrule
Ion 1 & 0.00669 & 0.01465 & 0.02363 & 0.03272 & 0.04106 & 0.04782 & 0.05232 & -0.05337 & 0.03733  \\
Ion 2 & -0.02273 & -0.04291 & -0.05424 & -0.05118 & -0.03324 & -0.00523 & 0.02439 & -0.04544 & 0.03733  \\
Ion 3 & 0.04046 & 0.05611 & 0.03493 & -0.01036 & -0.04684 & -0.04718 & -0.01107 & -0.03300 & 0.03733  \\
Ion 4 & -0.05373 & -0.03943 & 0.02271 & 0.05312 & 0.01267 & -0.04369 & -0.04005 & -0.01734 & 0.03733  \\
Ion 5 & 0.05863 & -0.00000 & -0.05404 & 0.00000 & 0.05272 & -0.00000 & -0.05117 & -0.00000 & 0.03733  \\
Ion 6 & -0.05373 & 0.03943 & 0.02271 & -0.05312 & 0.01267 & 0.04369 & -0.04005 & 0.01734 & 0.03733  \\
Ion 7 & 0.04046 & -0.05611 & 0.03493 & 0.01036 & -0.04684 & 0.04718 & -0.01107 & 0.03300 & 0.03733  \\
Ion 8 & -0.02273 & 0.04291 & -0.05424 & 0.05118 & -0.03324 & 0.00523 & 0.02439 & 0.04544 & 0.03733  \\
Ion 9 & 0.00669 & -0.01465 & 0.02363 & -0.03272 & 0.04106 & -0.04782 & 0.05232 & 0.05337 & 0.03733  \\
\bottomrule
\end{tabular}
\end{table*}

\begin{table*}
\caption{Lamb-Dicke parameters for eleven ions.}
\label{tab:eta11}
\scriptsize
\centering
\begin{tabular}{lrrrrrrrrrrr}
\toprule
 & Mode 1 & Mode 2 & Mode 3 & Mode 4 & Mode 5 & Mode 6 & Mode 7 & Mode 8 & Mode 9 & Mode 1 & Mode 11 \\
\midrule
Ion 1 & -0.00454 & 0.01003 & 0.01642 & -0.02324 & -0.03003 & -0.03634 & -0.04176 & -0.04589 & -0.04842 & 0.04851 & 0.03377  \\
Ion 2 & 0.01579 & -0.03161 & -0.04426 & 0.04980 & 0.04604 & 0.03305 & 0.01315 & -0.00956 & -0.03019 & 0.04348 & 0.03377  \\
Ion 3 & -0.02934 & 0.04841 & 0.04715 & -0.02367 & 0.01119 & 0.03969 & 0.04687 & 0.02888 & -0.00481 & 0.03545 & 0.03377  \\
Ion 4 & 0.04169 & -0.04896 & -0.01487 & -0.03163 & -0.04787 & -0.01891 & 0.02742 & 0.04684 & 0.02074 & 0.02504 & 0.03377  \\
Ion 5 & -0.05024 & 0.03066 & -0.02911 & 0.04582 & -0.00347 & -0.04691 & -0.02192 & 0.03436 & 0.03950 & 0.01295 & 0.03377  \\
Ion 6 & 0.05328 & 0.00000 & 0.04933 & -0.00000 & 0.04828 & 0.00000 & -0.04753 & 0.00000 & 0.04637 & -0.00000 & 0.03377  \\
Ion 7 & -0.05024 & -0.03066 & -0.02911 & -0.04582 & -0.00347 & 0.04691 & -0.02192 & -0.03436 & 0.03950 & -0.01295 & 0.03377  \\
Ion 8 & 0.04169 & 0.04896 & -0.01487 & 0.03163 & -0.04787 & 0.01891 & 0.02742 & -0.04684 & 0.02074 & -0.02504 & 0.03377  \\
Ion 9 & -0.02934 & -0.04841 & 0.04715 & 0.02367 & 0.01119 & -0.03969 & 0.04687 & -0.02888 & -0.00481 & -0.03545 & 0.03377  \\
Ion 10 & 0.01579 & 0.03161 & -0.04426 & -0.04980 & 0.04604 & -0.03305 & 0.01315 & 0.00956 & -0.03019 & -0.04348 & 0.03377  \\
Ion 11 & -0.00454 & -0.01003 & 0.01642 & 0.02324 & -0.03003 & 0.03634 & -0.04176 & 0.04589 & -0.04842 & -0.04851 & 0.03377  \\
\bottomrule
\end{tabular}
\end{table*}

\begin{table*}
\caption{Lamb-Dicke parameters for thirteen ions.}
\label{tab:eta13}
\scriptsize
\centering
\begin{tabular}{lrrrrrrrrrrrrr}
\toprule
 & Mode 1 & Mode 2 & Mode 3 & Mode 4 & Mode 5 & Mode 6 & Mode 7 & Mode 8 & Mode 9 & Mode 1 & Mode 11 & Mode 12 & Mode 13 \\
\midrule
Ion 1 & 0.00328 & -0.00728 & 0.01201 & -0.01721 & -0.02259 & -0.02790 & -0.03289 & -0.03731 & -0.04095 & -0.04359 & -0.04507 & 0.04473 & 0.03106  \\
Ion 2 & -0.01156 & 0.02391 & -0.03537 & 0.04335 & 0.04595 & 0.04220 & 0.03226 & 0.01742 & -0.00015 & -0.01771 & -0.03244 & 0.04129 & 0.03106  \\
Ion 3 & 0.02202 & -0.03984 & 0.04672 & -0.03820 & -0.01625 & 0.01141 & 0.03440 & 0.04385 & 0.03600 & 0.01387 & -0.01393 & 0.03577 & 0.03106  \\
Ion 4 & -0.03248 & 0.04715 & -0.03361 & -0.00147 & -0.03495 & -0.04387 & -0.02209 & 0.01493 & 0.04088 & 0.03730 & 0.00668 & 0.02847 & 0.03106  \\
Ion 5 & 0.04128 & -0.04189 & 0.00125 & 0.03955 & 0.03729 & -0.00523 & -0.04153 & -0.03210 & 0.01226 & 0.04255 & 0.02527 & 0.01979 & 0.03106  \\
Ion 6 & -0.04712 & 0.02458 & 0.03180 & -0.03896 & 0.01296 & 0.04442 & 0.00776 & -0.04028 & -0.02620 & 0.02773 & 0.03813 & 0.01014 & 0.03106  \\
Ion 7 & 0.04916 & -0.00000 & -0.04563 & -0.00000 & -0.04480 & 0.00000 & 0.04419 & 0.00000 & -0.04367 & -0.00000 & 0.04273 & 0.00000 & 0.03106  \\
Ion 8 & -0.04712 & -0.02458 & 0.03180 & 0.03896 & 0.01296 & -0.04442 & 0.00776 & 0.04028 & -0.02620 & -0.02773 & 0.03813 & -0.01014 & 0.03106  \\
Ion 9 & 0.04128 & 0.04189 & 0.00125 & -0.03955 & 0.03729 & 0.00523 & -0.04153 & 0.03210 & 0.01226 & -0.04255 & 0.02527 & -0.01979 & 0.03106  \\
Ion 10 & -0.03248 & -0.04715 & -0.03361 & 0.00147 & -0.03495 & 0.04387 & -0.02209 & -0.01493 & 0.04088 & -0.03730 & 0.00668 & -0.02847 & 0.03106  \\
Ion 11 & 0.02202 & 0.03984 & 0.04672 & 0.03820 & -0.01625 & -0.01141 & 0.03440 & -0.04385 & 0.03600 & -0.01387 & -0.01393 & -0.03577 & 0.03106  \\
Ion 12 & -0.01156 & -0.02391 & -0.03537 & -0.04335 & 0.04595 & -0.04220 & 0.03226 & -0.01742 & -0.00015 & 0.01771 & -0.03244 & -0.04129 & 0.03106  \\
Ion 13 & 0.00328 & 0.00728 & 0.01201 & 0.01721 & -0.02259 & 0.02790 & -0.03289 & 0.03731 & -0.04095 & 0.04359 & -0.04507 & -0.04473 & 0.03106  \\
\bottomrule
\end{tabular}
\end{table*}

\begin{table*}
\caption{Mode frequencies for seven ions.}
\label{tab:mode7}
\begin{tabular}{lrrrrrrr}
\toprule
 & Mode 1 & Mode 2 & Mode 3 & Mode 4 & Mode 5 & Mode 6 & Mode 7 \\
\midrule
Mode Frequencies & 2.70769 & 2.73603 & 2.77497 & 2.82075 & 2.86818 & 2.90951 & 2.93070  \\
\bottomrule
\end{tabular}
\end{table*}

\begin{table*}
\caption{Mode frequencies for nine ions.}
\label{tab:mode9}
\begin{tabular}{lrrrrrrrrr}
\toprule
 & Mode 1 & Mode 2 & Mode 3 & Mode 4 & Mode 5 & Mode 6 & Mode 7 & Mode 8 & Mode 9 \\
\midrule
Mode Frequencies & 2.70166 & 2.71968 & 2.74528 & 2.77700 & 2.81291 & 2.85055 & 2.88663 & 2.91626 & 2.93070  \\
\bottomrule
\end{tabular}
\end{table*}

\begin{table*}
\caption{Mode frequencies for eleven ions.}
\label{tab:mode11}
\begin{tabular}{lrrrrrrrrrrr}
\toprule
 & Mode 1 & Mode 2 & Mode 3 & Mode 4 & Mode 5 & Mode 6 & Mode 7 & Mode 8 & Mode 9 & Mode 1 & Mode 11 \\
\midrule
Mode Frequencies & 2.69850 & 2.71090 & 2.72882 & 2.75158 & 2.77827 & 2.80780 & 2.83879 & 2.86955 & 2.89779 & 2.92016 & 2.93070  \\
\bottomrule
\end{tabular}
\end{table*}

\begin{table*}
\caption{Mode frequencies for thirteen ions.}
\label{tab:mode13}
\begin{tabular}{lrrrrrrrrrrrrr}
\toprule
 & Mode 1 & Mode 2 & Mode 3 & Mode 4 & Mode 5 & Mode 6 & Mode 7 & Mode 8 & Mode 9 & Mode 1 & Mode 11 & Mode 12 & Mode 13 \\
\midrule
Mode Frequencies & 2.69664 & 2.70567 & 2.71886 & 2.73584 & 2.75612 & 2.77914 & 2.80420 & 2.83048 & 2.85695 & 2.88237 & 2.90508 & 2.92263 & 2.93070  \\
\bottomrule
\end{tabular}
\end{table*}

\section{Calibratability of parallel pulses}
\label{app:CaliIssues}
Quantum gate calibration is an important process where experimentally tunable parameters, such as laser power and frequency, of a gate pulse are adjusted so that, as an example, model-violation errors, arising from a mismatch between device-level physics and a simplified theoretical model, are reduced to result in a close-to-desired implementation of the target gate unitary in practice. In order to thus successfully run parallel gates in experiments, it is crucial that the gate pulses are tunable to admit calibrations. Recall now the graph pattern in Fig.~\ref{fig:CircuitsAndResultsHH}(a) for the HH simulation used as a benchmark in the main text, where the Hamiltonian was of the form $\sum_{i=1}^{4}\sigma^x_i\sigma_{i+1}^x$. This Hamiltonian implies a ring-graph of size four. In the case where the parallel gate pulses were solved at a qubit-by-qubit level (see \cite{grzesiak2020efficient}), i.e., the pulses seen by each of the four ions are solved to be $g_i(t)$, $i\in[1,5)$, the amplitude knobs $\Omega_i$ may be adjusted to result in the adjusted pulses $\Omega_i\cdot g_i(t)$. Suppose the entanglement angles obtained with $\Omega_i=1$ throughout are factors $\theta_{i,i+1}^{-1}$ away from the desired angles. To calibrate, we must now simultaneously satisfy the following system of equations:
\begin{align}
    \theta_{12}&=\Omega_1\Omega_2 \label{12} \\ 
    \theta_{23}&=\Omega_2\Omega_3 \label{23} \\
    \theta_{34}&=\Omega_3\Omega_4
    \label{34} \\
    \theta_{14}&=\Omega_1\Omega_4
    \label{14}
\end{align}
Setting $\theta_{12}=1$ without loss of generality and starting with (\ref{12}), we find
\begin{align}
\label{2_as_a_function_of_1}
    \Omega_2=\frac{1}{\Omega_1}.
\end{align}
Inserting (\ref{2_as_a_function_of_1}) into (\ref{23}), we obtain
\begin{align}
    \theta_{23}=\frac{\Omega_3}{\Omega_1} \rightarrow\Omega_3=\theta_{23}\Omega_1.
    \label{3_as_a_function_of_1}
\end{align}
Inserting (\ref{3_as_a_function_of_1}) into (\ref{34}), we obtain
\begin{align}
\theta_{34}=\theta_{23}\Omega_1\Omega_4\rightarrow\Omega_4=\frac{\theta_{34}}{\theta_{23}\Omega_1},
    \label{Const1}
\end{align}
which can be compared with (\ref{14}), rewritten as
\begin{align}
\label{Const2}
    \Omega_4=\frac{\theta_{14}}{\Omega_1}.
\end{align}
Eqs.~(\ref{Const1}) and (\ref{Const2}) cannot be simultaneously satisfied unless $\theta_{14}=\frac{\theta_{23}}{\theta_{34}}$.
In a practical system, such a constraint on errors is rarely satisfied, if at all, resulting in the method of~\cite{grzesiak2020efficient} being impractical to use in the above ring-graph pattern example considered, viewed through the lens of calibration. This is in a stark contrast to our method, where each parallel gate pulses are solved at a coupling-by-coupling level, i.e., ion $i$ is illuminated with $g_{i,i+1}(t)+g_{i-1,i}(t)$, where $g_{i,j}(t)$ denotes the pulse that implements an entangling gate between ions $i$ and $j$, affording us to control the coefficients $\Omega_{i,j}$'s independently for each gate to be calibrated, similar to a typical, single two-qubit gate calibration.

\end{document}